\documentclass[11pt,a4paper]{article} 
\pdfoutput=1
\usepackage{jheppub, hyperref}
\usepackage{array}
\newcolumntype{L}[1]{>{\raggedright\let\newline\\\arraybackslash\hspace{0pt}}m{#1}}
\newcolumntype{C}[1]{>{\centering\let\newline\\\arraybackslash\hspace{0pt}}m{#1}}
\newcolumntype{R}[1]{>{\raggedleft\let\newline\\\arraybackslash\hspace{0pt}}m{#1}}
\usepackage{float}
\usepackage{graphicx}
\usepackage{epsfig}
\usepackage{color}

\newcommand*{\be}{\begin{equation}}
\newcommand*{\ee}{\end{equation}}
\newcommand*{\bea}{\begin{eqnarray}}
\newcommand*{\eea}{\end{eqnarray}}

\newcommand{\comment}[1]{}

\newcommand{\cref}[1]{Chapter~\ref{c.#1}}

\def\beq{\begin{equation}}
\def\eeq{\end{equation}}
\def\bea{\begin{eqnarray}}
\def\eea{\end{eqnarray}}
\def\ba{\begin{array}}
	\def\ea{\end{array}}
\def\bi{\begin{itemize}}
	\def\ei{\end{itemize}}
\def\be{\begin{enumerate}}
	\def\ee{\end{enumerate}}
\def\bc{\begin{center}}
	\def\ec{\end{center}}
\def\bt{\begin{table}}
	\def\et{\end{table}}
\def\btb{\begin{tabular}}
	\def\etb{\end{tabular}}

	\def\lsim{\raise0.3ex\hbox{$\;<$\kern-0.75em\raise-1.1ex\hbox{$\sim\;$}}}
	\def\gsim{\raise0.3ex\hbox{$\;>$\kern-0.75em\raise-1.1ex\hbox{$\sim\;$}}}

\title{Gluon-Photon Signatures for color octet at the LHC (and beyond)}
\author[1]{G.Cacciapaglia,}
\author[1]{A.Deandrea,}
\author[2]{T. Flacke,}
\author[1]{A.M. Iyer}
\affiliation[1]{Univ. Lyon, Universit{\'e} Claude Bernard Lyon 1, CNRS/IN2P3, UMR5822 IP2I,\\ F-69622, Villeurbanne, France}
\affiliation[2]{Center for Theoretical Physics of the Universe, Institute for Basic Science (IBS), Daejeon 34126,
	Korea}
\abstract{
We consider a color octet scalar particle and its exotic decay in the channel gluon-$\gamma$ using an effective Lagrangian 
description for its strong and electromagnetic interactions. Such a state is present in many extensions of the Standard Model,
and in particular in composite Higgs models with top partial compositeness, where couplings to photons arise via the 
Wess-Zumino-Witten term. We find that final states with one or two photons allow for a better reach at the LHC, even for
small branching ratios. Masses up to $1.2$~TeV can be probed at the HL-LHC by use of all final states. Finally, we estimate 
the sensitivity of the hadronic FCC.}

\emailAdd{g.cacciapaglia@ipnl.in2p3.fr}
\emailAdd{deandrea@ipnl.in2p3.fr}
\emailAdd{flacke@ibs.re.kr}
\emailAdd{a.iyer@ipnl.in2p3.fr}

\begin{document}
\hspace*{112mm}{\large \tt CTPU-PTC-20-02} \\
\maketitle
\flushbottom

\section{Introduction}
Color octet  particles are present in various extensions of the Standard Model (SM), ranging from supersymmetric models to 
composite models for the electroweak sector. Examples include gluinos in supersymmetry, top-gluons 
in strong electroweak sectors, and Kaluza-Klein particles. Color octet properties have been widely discussed in the literature, with, in recent years, 
particular focus on the LHC physics, see for example \cite{Chen:2014haa} and references therein. This is justified by the huge potential for discovery or exclusion which the present and future options for the LHC offer in this specific sector.
In the following we shall focus on a particular class of color octet particles, those that are scalars or pseudo-scalars ($\Phi$), as 
they have a specific relevance for composite models for the electroweak sector \cite{Cacciapaglia:2015eqa}: composite color octets  
are typically made of fundamental fermions of an underlying strong dynamics associated to top partial compositeness \cite{Barnard:2013zea,Ferretti:2013kya}. 

Yet, the properties and strategies to determine bounds and future prospects for discovery do not depend crucially on the
specific model the color octet stems from. In fact, the couplings of the new state to SM particles are mainly dictated by their
gauge quantum numbers. First, thanks to QCD gauge interactions, the color octet scalar and pseudo-scalar can be pair-produced
at hadron colliders in a model independent way, with cross sections that only depend on the mass.
Single couplings to a pair of quarks, with typical preference for tops, are also allowed producing decays into a pair of jets or $t
\bar{t}$. Finally, loops of tops generate in turn couplings to a pair of gluons, to a gluon and a photon and to a gluon and a $Z$ boson.
In composite models, the loop induced couplings also receive a contribution from topological terms, i.e. the Wess-Zumino-Witten (WZW)
term. The coupling to gluons, and in minor extent the one to light quarks, also allows for single production. In composite scenarios, the 
WZW interactions are of particular interest as they carry information about the details
of the microscopic properties of the underlying dynamics, while the composite scalar and pseudo-scalar may be among the lightest
states of the theory if they arise as pseudo-Nambu-Goldstone bosons (pNGBs). A general analysis of jet-photon and jet-$Z$ resonances at the LHC
has been presented in \cite{Englert:2017bme}.

In this work, we will reconsider the phenomenology of a color octet scalar and pseudo-scalar by focusing on specific composite scenarios
with top partial compositeness. In fact, a common feature of models formulated in terms of a fermionic strongly coupled gauge theory \cite{Ferretti:2013kya}
 is the presence of specific additional (light) spin-0 resonances, namely two 
neutral singlets and a color octet pseudo-scalar \cite{Cacciapaglia:2015eqa,Belyaev:2016ftv}. 
In these models, the decay rate in the gluon-$\gamma$ channel can be predicted and turns out to be sizeable. Focusing
on this channel \cite{Hayot:1980gg,Belyaev:1999xe} is, therefore, particularly well motivated. 
We compare the gluon-gluon decay mode to the gluon-$\gamma$ one in the LHC setup. It is interesting to note that already in the 1980's 
these channels were compared at TeVatron \cite{Hayot:1980gg} for their potential in the search of a strongly interacting electroweak 
sector. We discuss the implication and the potential of these modes for the color octet $\Phi$ at the LHC and its future high 
luminosity (HL-LHC) and high energy (HE-LHC) options, as well as future projects (FCC-hh). This will allow to define the detailed analysis strategies for the experimental searches at the LHC, and at future options, of these kinds of resonances.  Our work is of particular interest in view of testing models with a strong electroweak sector. 
 

In order to discuss in a general way the color octet interactions across different models, we shall consider effective interactions encoded in the effective Lagrangian for a pseudo-scalar octet discussed in \cite{Belyaev:2016ftv}, which contains general features present in typical extensions of the SM. In particular, the color octet decays into $t\bar{t}$, $gg$, $g\gamma$, and $gZ$ are parameterized as follows:
\bea
\mathcal{L}_{\Phi} &=& \frac{1}{2} (D_\mu \Phi^a)^2- \frac{1}{2} M_{\Phi}^2 (\Phi^a)^2 + i\ C_{t} \frac{m_t}{f_{\Phi}} \Phi^a\ \bar{t}\gamma_5 \frac{\lambda^a}{2} t  \nonumber \\&& + \frac{\alpha_s \kappa_{g}}{8 \pi f_{\Phi}}\Phi^a\ \epsilon^{\mu\nu\rho\sigma}\left[ \frac{1}{2}d^{abc}\ G^b_{\mu\nu}G^c_{\rho\sigma}+\frac{{e} \kappa_{\gamma}}{{g_s}  \kappa_{g}}\ G^a_{\mu\nu}F_{\rho\sigma}-\frac{{e} \tan \theta_W \kappa_{Z}}{{g_s}  \kappa_{g}}\ G^a_{\mu\nu}Z_{\rho\sigma}  \right]\,,
\label{eq:Lag2}
\eea
where $f_\Phi$ is a mass scale (corresponding to the decay constant of the composite $\Phi$), while the covariant derivative contains QCD interactions with gluons. The relative value of the photon coupling, $\kappa_\gamma$, and the $Z$ coupling, $\kappa_Z$, depend on the electroweak quantum numbers of the multiplet $\Phi$ belongs to. In the following, for simplicity, we will focus on a weak isosinglet, for which
\begin{equation}
\kappa_\gamma = \kappa_Z \equiv \kappa_B\,,
\end{equation}
as this case applies directly to composite Higgs examples.
As a bookkeeping, we present other cases in Appendix \ref{app:bookkeeping}.
In the underlying models considered in \cite{Belyaev:2016ftv}, the color octet arises as a bound state of color triplet fermions $\chi$ with hypercharge $Y_\chi =1/3$ or $2/3$, thus the ratio $\kappa_B/\kappa_g = 2 Y_\chi$ is also fixed. In turn, this property fixes the relative branching fractions amongst the bosonic final states, as given in Table \ref{tab:octetBR}. We will use these branching fractions as benchmarks, but results will be presented also for generic $\kappa_B/\kappa_g$.
\begin{table}[t]
\begin{center}
			\begin{tabular}{|c|c|c|}
			\hline
			& $ \frac{ {\mathrm{BR}}(\Phi\to g\gamma)}{{\mathrm{BR}}(\Phi\to gg)}  $ & $ \frac{ {\mathrm{BR}}(\Phi\to g Z)}{{\mathrm{BR}}(\Phi\to gg)}$  \\
			\hline\hline
			$Y_\chi=  1/3$    &  $0.048$   & $0.014$  \\
			$Y_\chi= 2/3 $    &   $0.19$  &   $0.058$  \\
			\hline
	\end{tabular}
\end{center}
	\caption{Values of ratios of BRs in di-bosons for the pseudo-scalar octet isosinglet at a mass of $1$~TeV. The mass
		fixes the dependence due to the running of the strong gauge coupling,
		$\alpha_s(1~{\rm TeV}) = 0.0881$ is used for this evaluation. The $Y_\chi=1/3\ (2/3)$ will be referred to as the pessimistic (optimistic) case corresponding to its reach in the photon channels, while the decay into $gZ$ is always subleading.}
	\label{tab:octetBR}
\end{table}

 The ratio of the partial widths to tops vs. gluons is  \cite{Belyaev:2016ftv}
\begin{equation}
\frac{\Gamma_{\Phi\rightarrow t\bar{t}}}{\Gamma_{\Phi\rightarrow gg}} =\frac{48\pi^2}{5\alpha_s^2} \frac{C_t^2}{\kappa_g^2}\frac{m_t^2}{M^2_\Phi} \left(1-4\frac{m_t^2}{M^2_\Phi}\right)^{1/2}\,,
\end{equation}
thus it scales with the ratio $C^2_t/\kappa^2_g$, which we leave as a free parameter. Note that all ratios of branching fractions are independent on the scale $f_\Phi$, which is only relevant for the total width of the color octet (and single production rates). In the models we consider, the total width is always very small compared to the mass.

\section{Current bounds on color octet single and pair production}

\begin{figure}[t]
	\begin{center}
		\begin{tabular}{cc}
			\includegraphics[width=7.2cm]{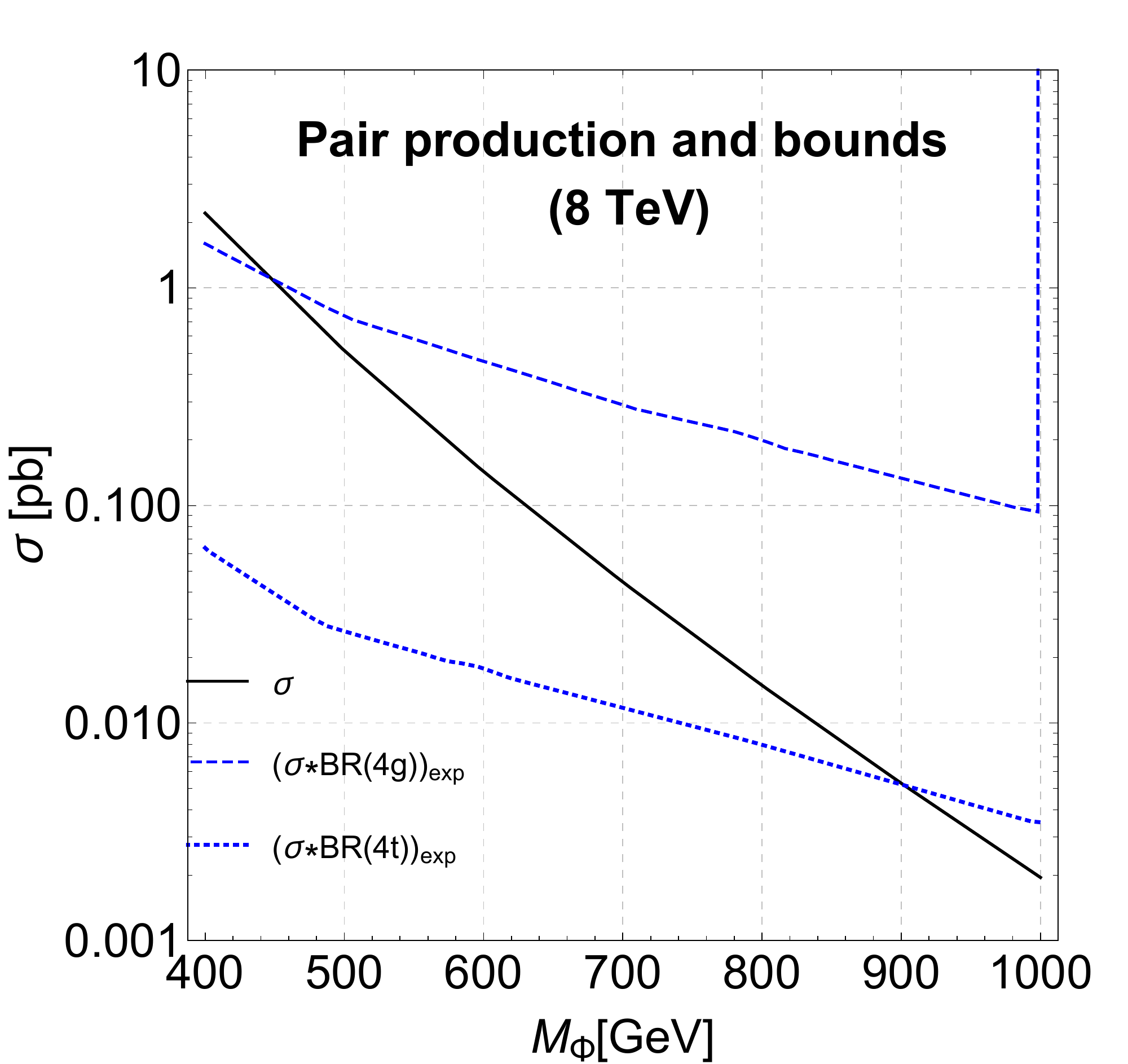}&\includegraphics[width=7.2cm]{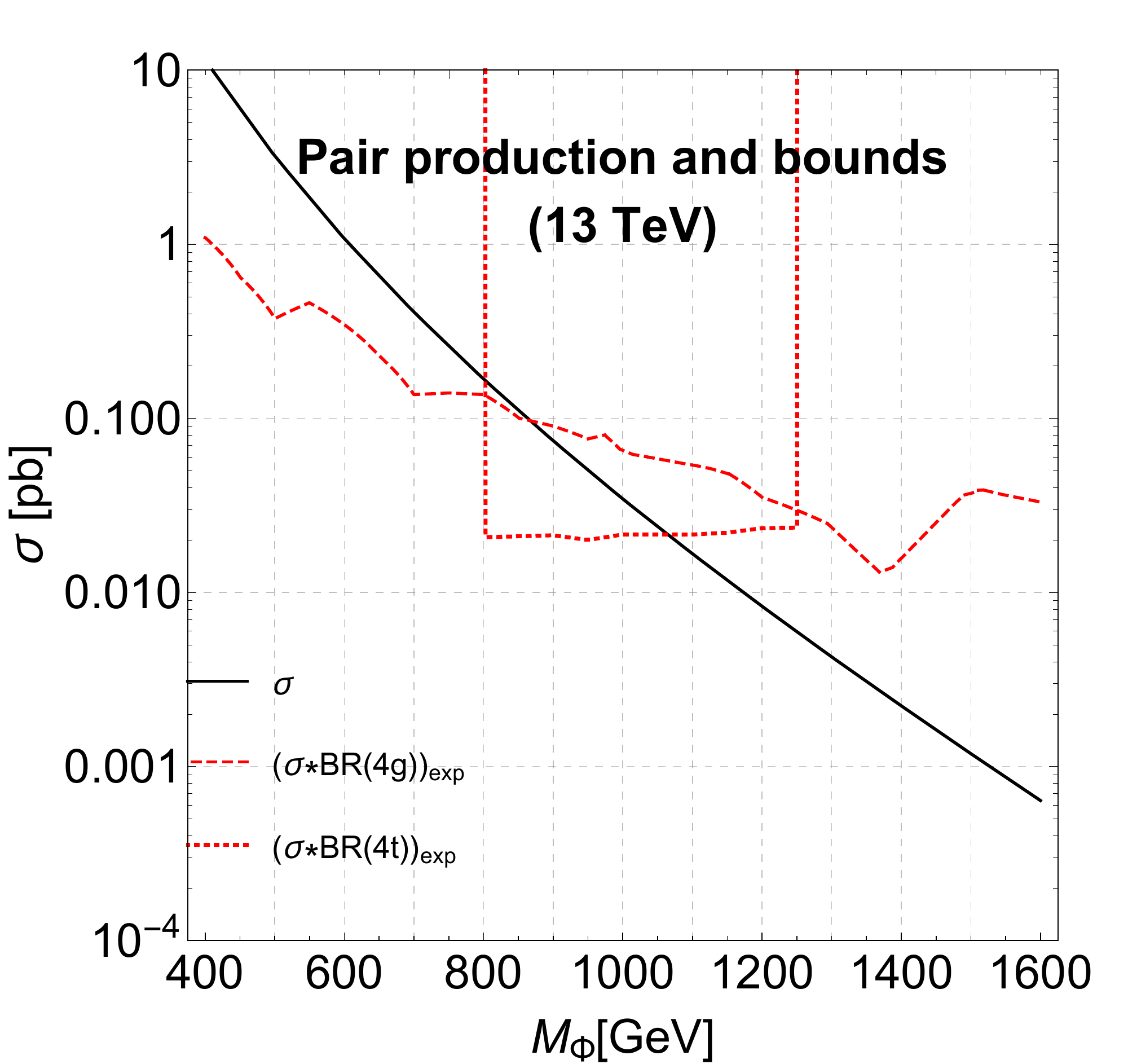}		
		\end{tabular}
	\end{center}
	\caption{Observed LHC bounds on cross sections from 4-top searches and di-jet-pair searches at run I (left) and run II (right). For reference, the black lines show the  total QCD pair production cross section at the respective center-of-mass energy.}
	\label{fig:pairexpbds}
\end{figure}

\begin{figure}[t]
	\begin{center}
		\begin{tabular}{cc}
			\includegraphics[width=7.2cm]{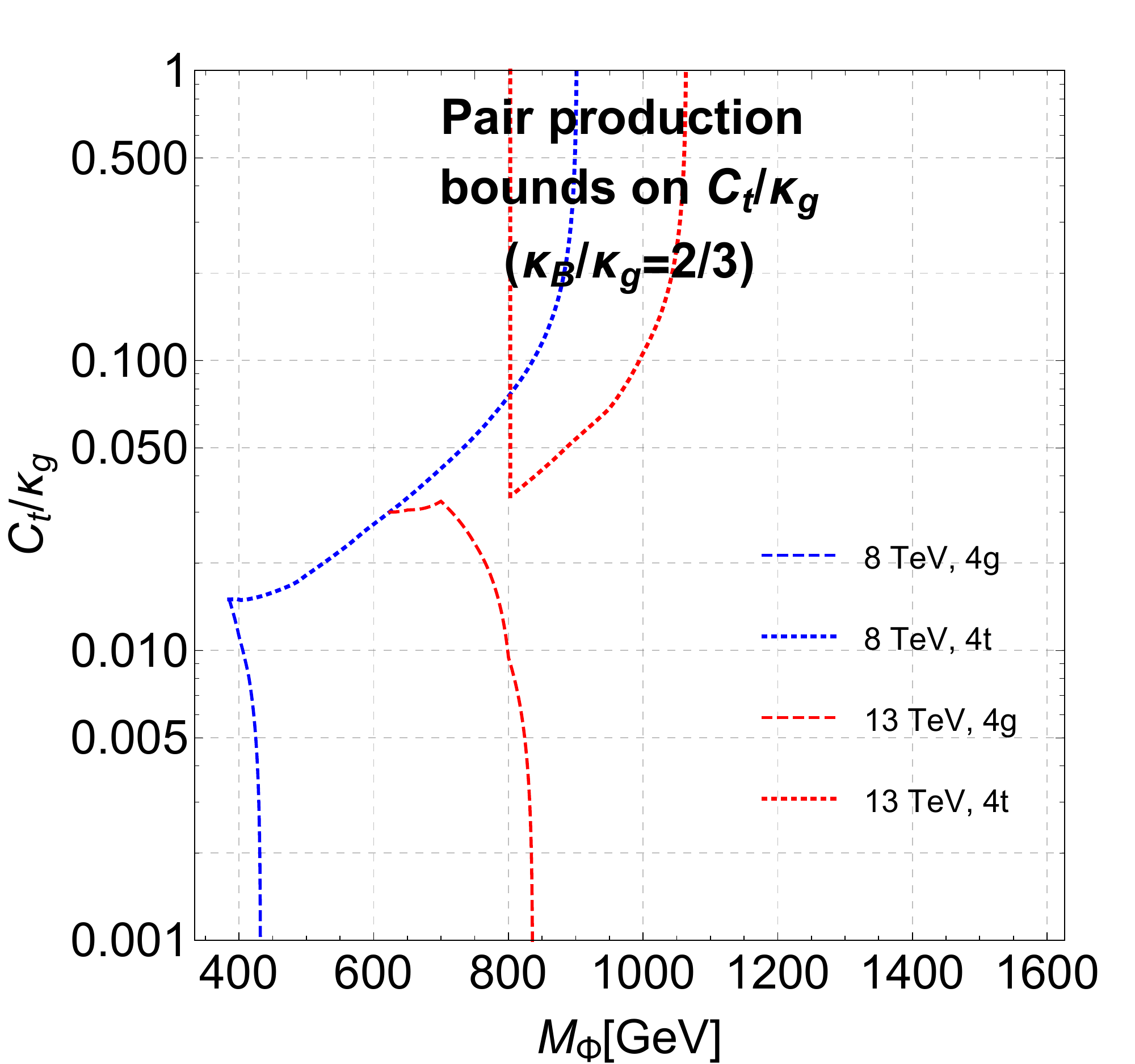}&\includegraphics[width=7.2cm]{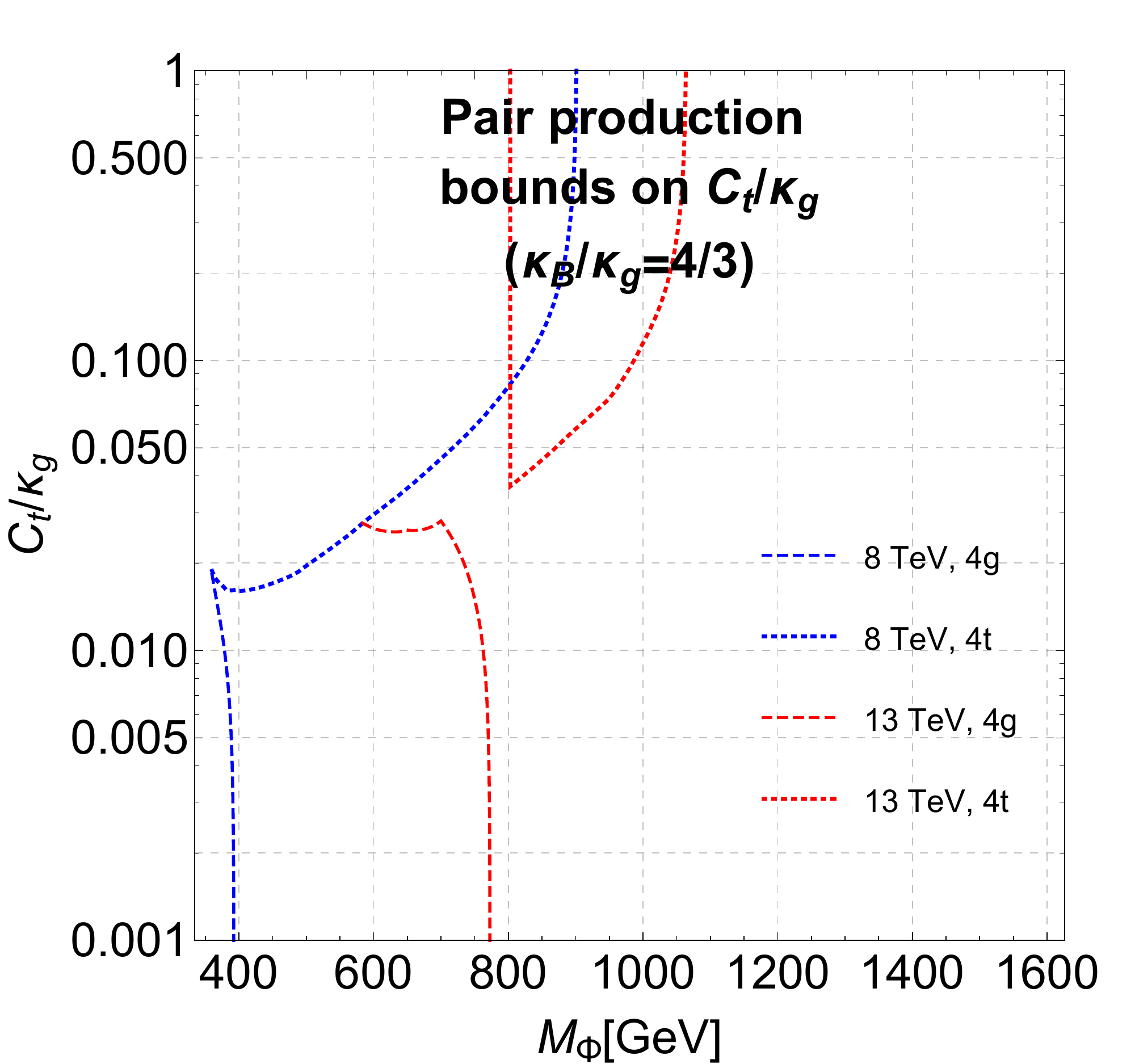}		
		\end{tabular}
	\end{center}
	\caption{Excluded regions in the $M_\Phi$ vs. $C_t/\kappa_g$ parameter space for the two benchmark models, with branching ratios given in Table \ref{tab:octetBR}.}
	\label{fig:pairpsbds}
\end{figure}

For QCD pair production of $\Phi$, with subsequent decays into two pairs of $t\bar{t}$ or two pairs of gluons, existing 4-top searches and searches for a pair of di-jet 
resonances yield bounds on the mass of the color octet that only depend on the branching ratios $\mathrm{BR}(\Phi\to t\bar{t})$ and $\mathrm{BR}(\Phi\to gg)$ (for the benchmark composite models, only the ratio $C_t/\kappa_g$ is relevant, as the relative rates in $g\gamma$ and $gZ$ are fixed). 
Searches for 4-top final states \cite{Aad:2015gdg,Aad:2015kqa}  at the LHC run I were interpreted in a color octet model (sgluon), thus they can be directly applied to our case, while searches for di-jet pairs are not very sensitive to the color structure of the decaying resonances. 
In Figure \ref{fig:pairexpbds} (left) we show the run I bounds on the cross section for the above-mentioned 4-top searches \cite{Aad:2015gdg,Aad:2015kqa} and for the jet final state \cite{Khachatryan:2014lpa}. The solid black line shows, for reference, the QCD pair production at  $\sqrt{s}=8$~TeV at LO in QCD\footnote{We calculate the  color octet pair production cross section at leading order using MadGraph 5  with the NNPDF23LO (as\_0130\_qed) PDF set without applying any $K-$ factor. As shown in \cite{Degrande:2014sta}, the NLO $K-$ factor for color octet pair production is close to one for $M_\Phi$ at the TeV scale.}.
At run II, the color octet interpretation for 4-top searches has been dismissed, thus we need to use a recast of the searches, which is only available in Ref.~\cite{Darme:2018dvz} for the same-sign lepton search of Ref.\cite{Sirunyan:2017roi}\footnote{The CMS 4-top search in the same-sign lepton channel \cite{Darme:2018dvz} is based on the 36 fb$^{-1}$ dataset. A CMS search with 137 fb$^{-1}$ became available recently \cite{Sirunyan:2019wxt}. Further 4-top ATLAS and CMS searches with 36  fb$^{-1}$ are also available \cite{Aaboud:2018xuw,Aaboud:2018xpj,Aaboud:2018jsj,Sirunyan:2019nxl} but require non-trivial recasting in order to obtain a bound on color octet resonances. We therefore restrict ourselves to  \cite{Sirunyan:2017roi} for which the recast  \cite{Darme:2018dvz} is available.}, which is based on an integrated luminosity of $35.9~\mbox{fb}^{-1}$. In Figure \ref{fig:pairexpbds} (right) we show the bound on the cross section, together with the ATLAS and CMS jet searches \cite{Aaboud:2017nmi,Sirunyan:2018rlj} that are based on $36.7~\mbox{fb}^{-1}$ and $35.9~\mbox{fb}^{-1}$ integrated luminosity respectively, together with the LO QCD cross section at $\sqrt{s}=13$~TeV.
To translate these bounds into a limit on the color octet mass, it is enough to rescale the total production cross section by the branching ratios, which only depend on ratios of couplings.
For the two benchmark models, with reference values $\kappa_B/\kappa_g = 2/3$ and $4/3$, we show the excluded regions in the $M_\Phi$ vs. $C_t/\kappa_g$ plane in Figure \ref{fig:pairpsbds}. The bounds for $\kappa_B/\kappa_g=4/3$ are marginally weaker because branching fractions into $g\gamma$ (and $gZ$) are larger in this case, and events with these decays evade detection in the 4-top and di-jet-pair searches. We see that the bounds on $M_\Phi$ range from $\sim 800$~GeV in the 4-jet region to $\sim 1$~TeV in the 4-top region, with a `hole' reaching down to $\sim 600$~GeV for intermediate $C_t/\kappa_g \approx 5\%$ due to the run II 4-top search loosing steam because of triggers.

\begin{figure}[t]
 	\begin{center}
 		\begin{tabular}{cc}
 			\includegraphics[width=7.2cm]{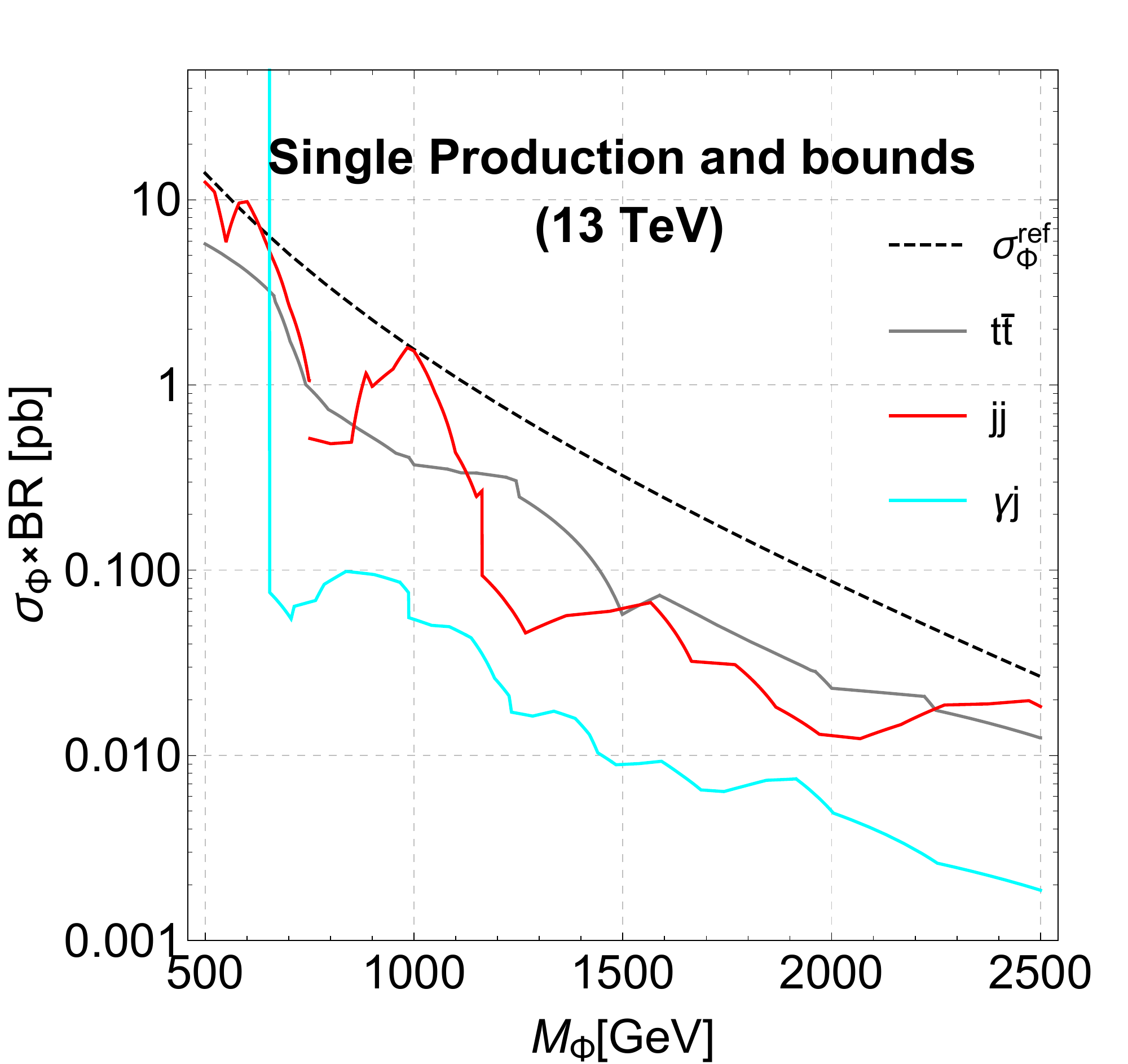}&\includegraphics[width=7.2cm]{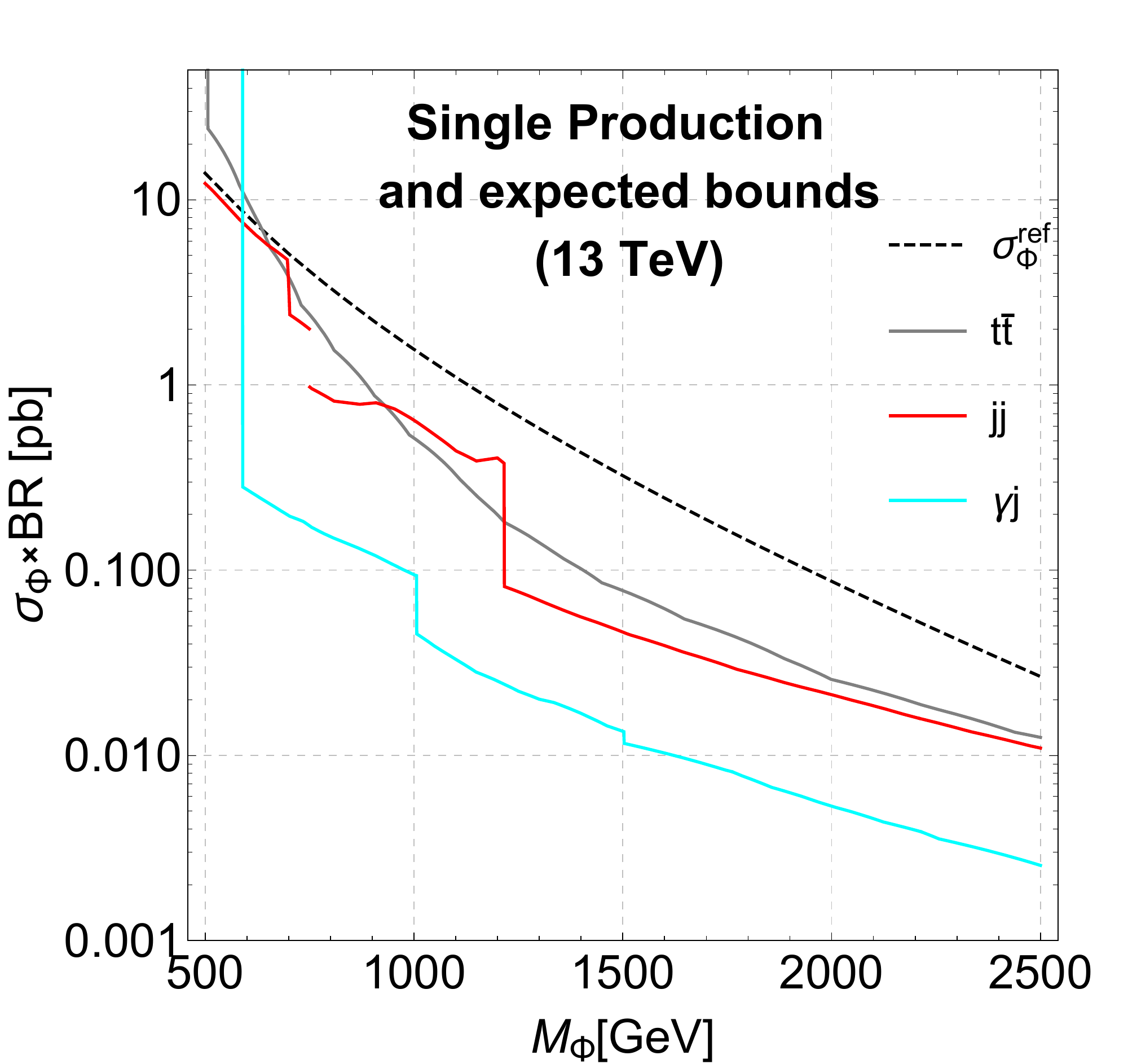}
 		\end{tabular}
 	\end{center}
 	\caption{Current observed (left) and expected (right) experimental bounds on the production cross section of $t\bar{t}$, $jj$, and $j\gamma$ resonances as a function of the resonance mass. As a reference, we give the single production cross section of $\Phi$ for $\kappa_g / f_\Phi = 10 \mbox{ TeV}^{-1}$ as a function of $M_\Phi$. The production cross section scales with $(\kappa_g / f_\Phi)^2 $.}
 	\label{fig:singlepsbds}
 \end{figure}

The color octet $\Phi$ can also be singly produced in gluon fusion via its WZW interaction (top loops give a sub-leading contribution in the relevant mass range that is not excluded by pair production). Unlike QCD pair production, the single production cross section depends not only on the octet mass but also on $\kappa_g/f_\Phi$.\footnote{In models with BSM QCD bound state color octets, further single-production mechanisms are possible, see e.g. \cite{Bhattacherjee:2017cxh}.} The different resonant final states are $t\bar{t}$, $gg$, $g\gamma$, and $gZ$, where the branching fraction between the $t\bar{t}$ and the gauge boson final states is controlled by $C_t/\kappa_g$, while the ratios between the boson channels with $\gamma$/$Z$ and purely hadronic are controlled by $\kappa_B/\kappa_g$ (benchmark values given in Table~\ref{tab:octetBR}). These final states are covered by run II resonance searches in $t\bar{t}$ \cite{Sirunyan:2018ryr,Aaboud:2018mjh,Aaboud:2019roo} with $36~\mbox{fb}^{-1}$ integrated luminosity,  low-mass and high-mass $jj$ searches \cite{Sirunyan:2018xlo,CMS:2018wxx,Aaboud:2017yvp,Aaboud:2018fzt,Aad:2019hjw} with $36$--$139~\mbox{fb}^{-1}$ datasets, and the excited quark searches in $j\gamma$ \cite{Aaboud:2017nak,Sirunyan:2017fho} at $\sqrt{s}= 13$~TeV based on $\sim 36~\mbox{fb}^{-1}$, while no direct search is available for a $jZ$ resonance (which, however, has a low branching ratio in our focus models). 
 For the $j\gamma$ final state, the $13$~TeV searches only apply to invariant masses above $1$~TeV, therefore we also consider run I searches at $8$~TeV \cite{Aad:2013cva,Khachatryan:2014aka} to cover the lower mass range.
 In Figure~\ref{fig:singlepsbds} (left) we collect the observed bounds on cross section times branching ratio in the various channels, together with the single $\Phi$ cross section for $\kappa_g / f_\Phi = 10 \mbox{ TeV}^{-1}$ for reference. For the $j\gamma$ searches at $8$~TeV, the observed bound (relevant for $600 < M_\Phi < 1000$~GeV), is plotted rescaled by the ratio of production cross section in the two energy regimes for the color octet, i.e. we plot $\sigma_{\rm limit} (j\gamma) \times \sigma_{13~{\rm TeV}} (gg \to \Phi)/\sigma_{8~{\rm TeV}} (gg \to \Phi)$.

 The cross section limits can be directly translated into bounds on the parameter space of the color octet. In particular, for fixed branching ratios, we can extract an upper bound on the couplings to gluon, $\kappa_g/f_\Phi$, relevant for the rates in single production.
Some plots visualizing the observed bounds can be found in Appendix~\ref{app:bounds}. Here, however, we want to focus on another point: comparing the reach of the di-jet, $t\bar{t}$ and jet-$\gamma$, we would like to highlight in what parameter space the decay model with a photon becomes relevant, notwithstanding the smaller branching ratio. As the observed bounds strongly vary with $M_\Phi$ due to statistical fluctuations, see left panel of Figure~\ref{fig:singlepsbds}, to obtain more clear indications we used the expected bounds, as shown in the right panel of Figure~\ref{fig:singlepsbds}. 
 
 	\begin{figure}[t]
	\begin{center}
		\begin{tabular}{cc}
			\includegraphics[width=7.2cm]{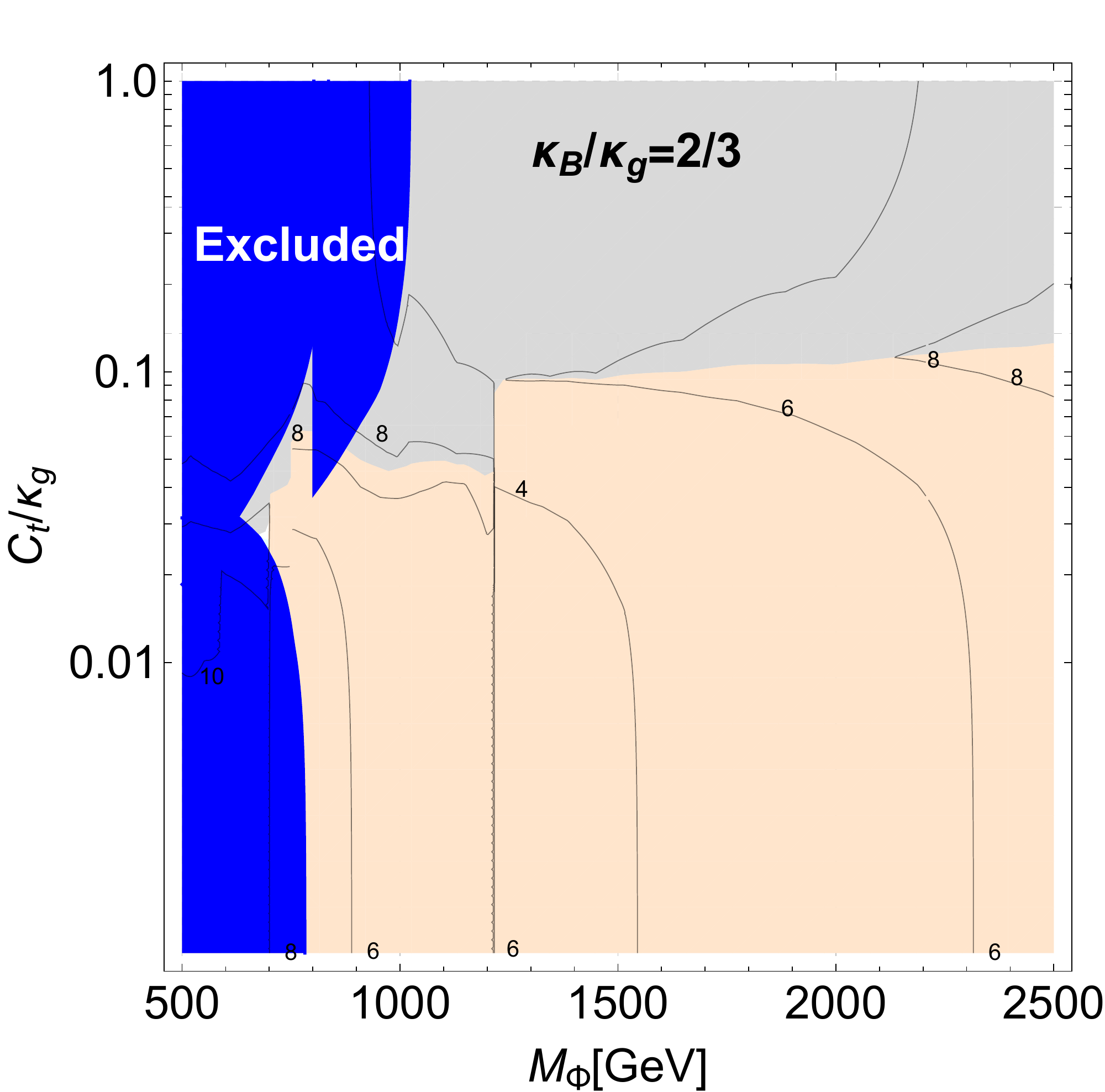}&\includegraphics[width=7.2cm]{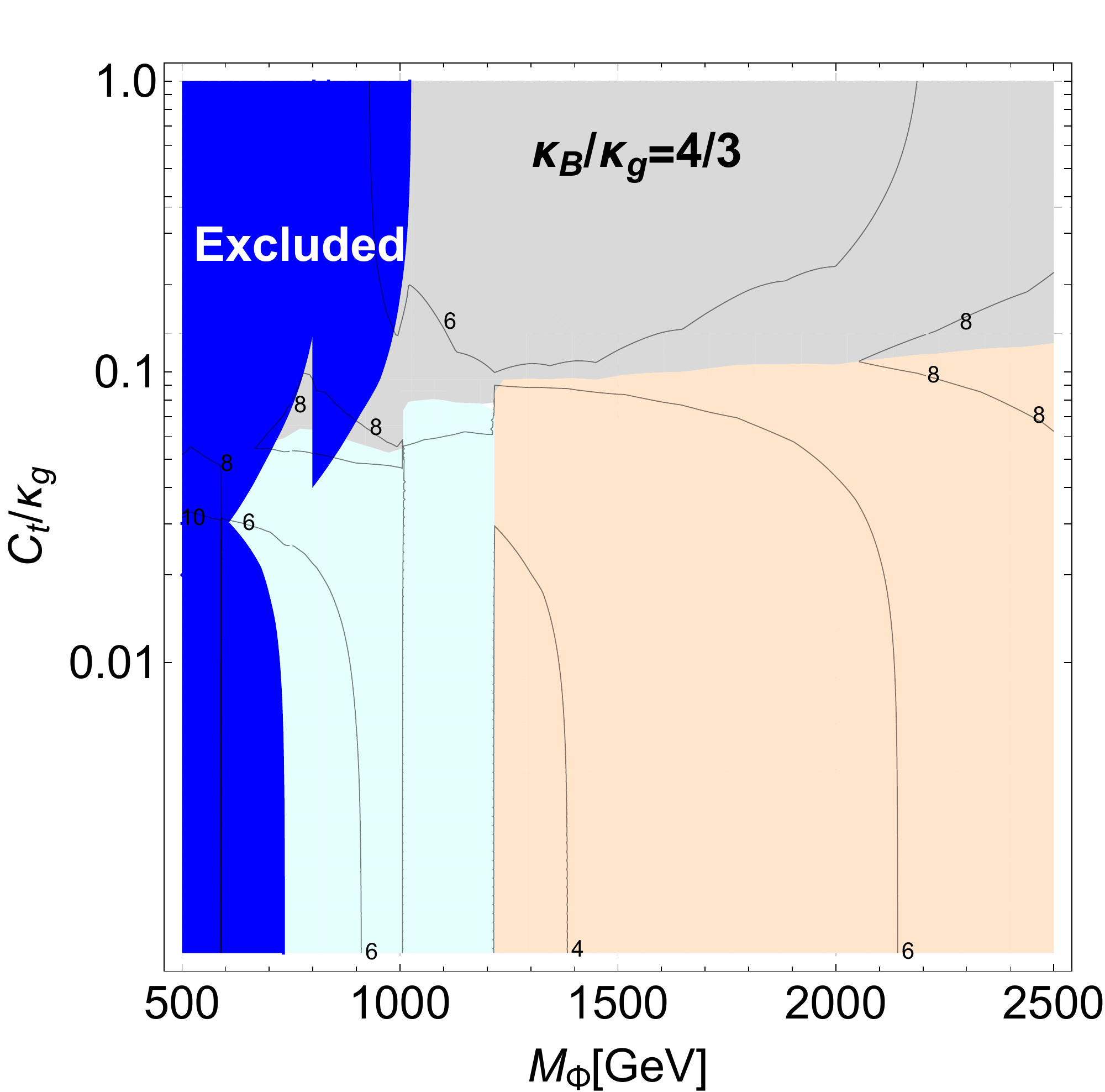}		
		\end{tabular}
	\end{center}
	\caption{Expected bounds on color octet single production in the $M_\Phi$ vs. $C_t/\kappa_g$ plane for  $\kappa_B/\kappa_g = 2/3$ and $4/3$. The contours show the  upper bound on $\kappa_g/f_\Phi$ (in TeV$^{-1}$). The dark-blue areas are excluded by pair production searches. In the  grey areas, the strongest bound arises from $t\bar{t}$ resonance searches. In the orange (cyan) areas, the currently strongest bound arises from di-jet (jet-$\gamma$) searches.}
	\label{fig:singlepsbds2}
\end{figure}

 \subsection{On the relevance of the photon} \label{sec:photon}
 
 As in general there are too many free parameters, we first focus on the two benchmark models: in Fig.~\ref{fig:singlepsbds2} we show the upper limit on $(\kappa_g / f_\Phi)$ in the plane $C_t/\kappa_g$ vs. $M_\Phi$, where the dark blue region is excluded by pair production (C.f. Fig.~\ref{fig:pairpsbds}). As already mentioned, we use the expected bounds in order to obtain more readable figures, where the actual observed bound is numerically close up to statistical fluctuations (see Fig.~\ref{fig:singlepsbds} and plots in Appendix~\ref{app:bounds}).
The shaded regions indicate which channel provides the strongest limit, with grey corresponding to $t\bar{t}$, orange to $jj$ and cyan to $j\gamma$. The comparison is, however, not completely fair because some $jj$ searches include $139~\mbox{fb}^{-1}$ integrated luminosity while $t\bar{t}$ and $j\gamma$ are based on only $36~\mbox{fb}^{-1}$ of data (with the latter also partially based on $8$~TeV searches). Thus the grey and cyan regions underestimate the actual potential of these two final states. Nevertheless, the plots clearly show that $t\bar{t}$ dominates as soon as a significant coupling to tops is present, $C_t/\kappa_g \gtrapprox 0.1$.
 
 	\begin{figure}[t]
	\begin{center}
		\begin{tabular}{c}
			\includegraphics[width=8cm]{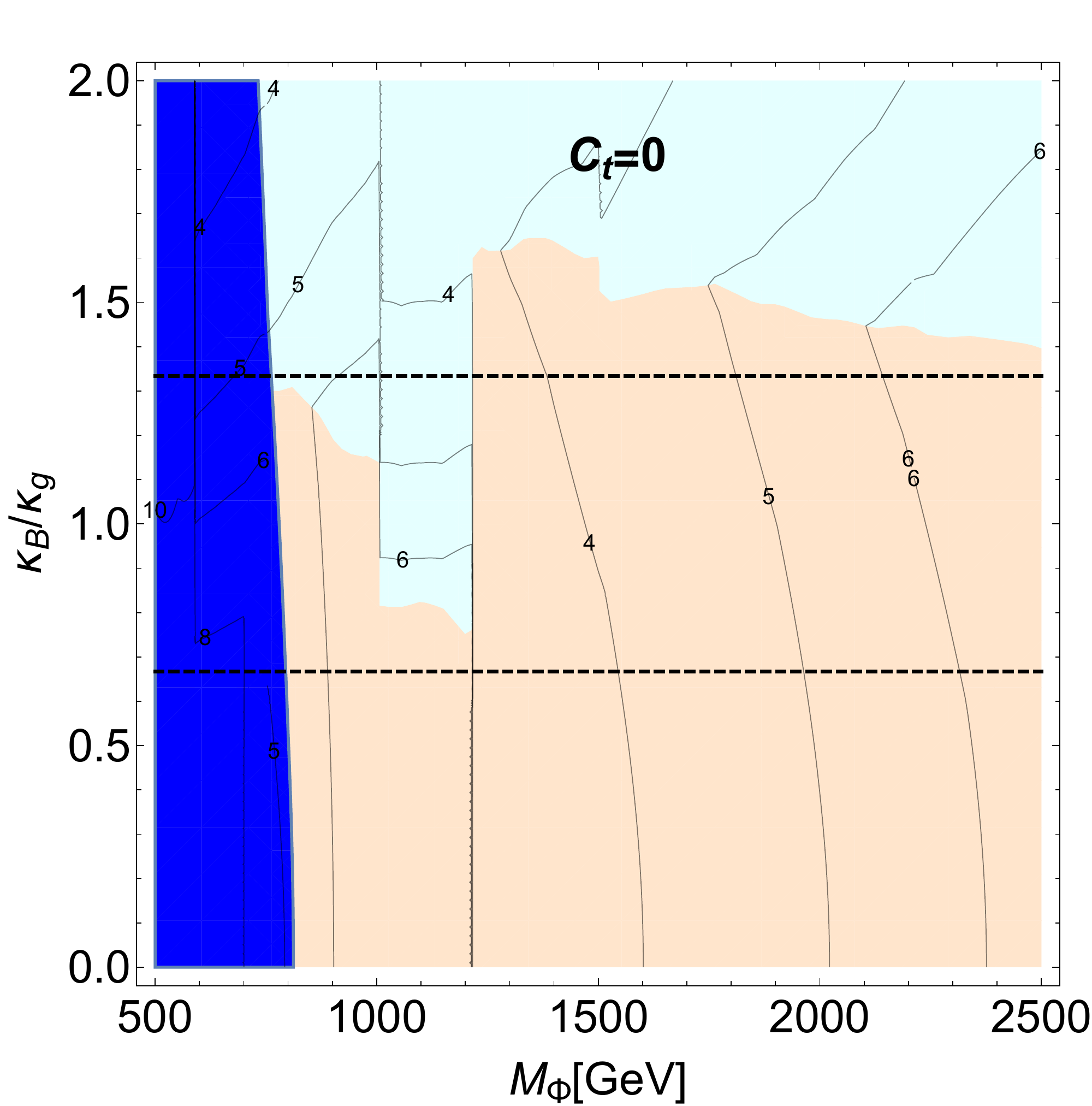}		
		\end{tabular}
	\end{center}
	\caption{Expected bounds on color octet single production in the $M_\Phi$ vs. $\kappa_B/\kappa_g$ plane. The dark blue area is excluded from pair production bounds, while the orange (cyan) area labels where the dominant bound arises from the $jj$ ($j\gamma$) searches. 
		Contours show the upper bound on $\kappa_g/f_\Phi$ in TeV$^{-1}$. For reference, the horizontal dashed lines indicate fixed $\kappa_B/\kappa_g$ ratios of $2/3$ (lower) and $4/3$ (higher). }
	\label{fig:singlepsbdsYchi}
\end{figure}

To show more general results, not only limited to the benchmark models, we now focus on the region where the $t\bar{t}$ final state is negligible and the decays into gauge bosons dominate. A critical discussion of the interest of this assumption in composite Higgs models of reference is presented in the next subsection.
 In Fig.~\ref{fig:singlepsbdsYchi} we show bounds on $\kappa_g/f_\Phi$ in the plane $\kappa_B/\kappa_g$ vs. $M_\Phi$ for $C_t=0$, with the same conventions for the shaded regions as in Fig.~\ref{fig:singlepsbds2}.  The horizontal dashed lines correspond to the two benchmark models, showing that the $Y_\chi=1/3$ bounds are currently dominated by the $jj$ final state (with the caveat on the lower luminosity/energy available in the $j\gamma$ channel), while for $Y_\chi=2/3$, bounds from the $jj$ and $j\gamma$ final state compete. Inclusion of more data in the $j\gamma$ channel will, however, further extend the relevance of the $j\gamma$ final state compared to the current plot. 
These results clearly show that the final state with a photon is typically very important in this class of models, and it will be our main focus in the next section.

\subsection{The bearable smallness of $C_t$ in composite models}

From the single and pair production bounds we see that the bosonic decay modes (into di-jets or jet-photon) only play a dominant role if $C_t/\kappa_g\lesssim 0.1$, as otherwise the 4t and $t\bar{t}$ searches yield best bounds. A priori, the WZW couplings of the color octet $\Phi$ to gauge bosons and its couplings to SM fermions are not directly related, such that small or even vanishing $C_t/\kappa_g\lesssim 0.1$ is viable. Nevertheless, in models where $\Phi$ arises from a composite sector which is the source of electroweak symmetry breaking and the generation of the top-quark mass, the underlying structure can link WZW and fermion couplings. In this section, we therefore review the relation between top and WZW couplings in composite Higgs models with top partial compositeness in order to assess whether a small $C_t$ can be naturally obtained. We remark that these models are one of many examples to which our analysis can apply,  so they should just be taken as a guide.

The models of top partial compositeness we consider were first proposed in~\cite{Barnard:2013zea,Ferretti:2013kya}: they consist in a confining gauge group $\mathcal{G}_{\rm HC}$ with two species of fermions, $\psi$ and $\chi$, which transform under different irreducible representations of $\mathcal{G}_{\rm HC}$. In particular, the $\psi$'s carry only electroweak quantum numbers and are responsible for generating a composite Higgs upon condensation. The $\chi$'s carry QCD charges and hypercharge and are introduced to obtain fermionic bound states which serve as vector-like top partners (in the following called ''baryons''). The quantum numbers  are chosen so that baryons, i.e. spin-1/2 resonances, are formed out of the two species: depending on the hypercharge $Y_\chi = 1/3$ or $2/3$, we have:
\beq
\mathcal{B}_j = \left\{ \begin{array}{l}
\phantom{\Big(} \langle \psi \psi \chi \rangle\;\; \mbox{or}\;\; \langle \psi \bar{\psi} \bar{\chi} \rangle\,, \quad \mbox{for} \;\; Y_\chi = 2/3\,; \\
\phantom{\Big(} \langle \psi \chi \chi \rangle\;\; \mbox{or}\;\; \langle \bar{\psi} \bar{\chi} \chi \rangle\;\; \mbox{or}\;\; \langle \psi \bar{\chi} \bar{\chi} \rangle\,, \quad \mbox{for} \;\; Y_\chi = 1/3\,. 
\end{array} \right. \label{eq:baryons1}
\eeq
The baryons $\mathcal{B}_j$ are the resonances that mix to the SM top fields in order to give them mass via the composite Higgs.

Beyond the top-partners, there are further bound states made up from $\chi$'s. In particular, analogous to the Higgs from $\psi$ condensation, the $\chi$'s  generate a color octet pNGB $\Phi$.  \footnote{The composite Higgs (by construction) and the color octet are present in all models of top partial compositeness. The models also contain further neutral, electroweakly charged, and colored pNGBs, which depend on the number and representations of $\chi$'s and $\psi$'s under $\mathcal{G}_{\rm HC}$ (see \cite{Ferretti:2013kya,Belyaev:2016ftv,Cacciapaglia:2020vyf}).} 
The WZW couplings of the color octet bound state $\Phi$ are determined by the quantum numbers of the underlying $\chi$,  while the coupling to tops depends on the nature of the baryons $\mathcal{B}_j$ coupling to the left and right-handed tops.
In Ref.~\cite{Belyaev:2016ftv}, it was found that
\beq
\frac{C_t}{\kappa_g} = \frac{n_\chi}{d_\chi}\,, 
\label{eq:ctokg1}
\eeq
where $d_\chi$ is the dimension of the underlying $\chi$ fermion under the confining gauge group, while $n_\chi = n_{\chi L} + n_{\chi R}$ is an integer, where $n_{\chi L/R}$ count how many $\chi - \bar{\chi}$ fermions appear in the baryons mixing to left and right-handed tops respectively.  From Eq.~\eqref{eq:baryons1}, we see that this depends on the hypercharge of $\chi$. For $Y_\chi = 2/3$, $n_{L/R} = \pm 1$, and thus $n_\chi = \pm 2, 0$. For $Y_\chi = 1/3$, $n_{L/R} = \pm 2, 0$, and thus $n_\chi = \pm 4, \pm 2, 0$.
In the case $n_\chi = 0$, therefore, vanishing couplings to the top can be obtained. Furthermore, $C_t \ll \kappa_g$ could also be realised in models where $d_\chi$ is sufficiently large. 
In Table~\ref{tab:M1_12} we list values of $d_\chi$  for the  12 minimal models of top partial compositeness of \cite{Ferretti:2013kya,Belyaev:2016ftv,Cacciapaglia:2020vyf}. We see that while for instance M2 and M7 have moderately large $d_\chi$, achieving $C_t/\kappa_g < 0.1$ is only possible  within the minimal models with $n_\chi = 0$.

\begin{table}[h]
\begin{center}
			\begin{tabular}{|l|c|c|c|c|c|c|}
			\hline
			 & M1 & M2 & M3 & M4 & M5 & M6 \\
			\hline
			$\mathcal{G}_{\rm HC}$ & SO(7) & SO(9) & SO(7) & SO(9) & Sp(4) & SU(4) \\
			\hline
			$Y_\chi$ & $1/3$ & $1/3$ & $2/3$ & $2/3$ & $1/3$ & $1/3$ \\
			$d_\chi$ & $8$ & $16$ & $7$ & $9$ & $4$ & $4$ \\
			\hline \hline
			& M7 & M8 & M9 & M10 & M11 & M12 \\
			\hline
			$\mathcal{G}_{\rm HC}$ & SO(10) & Sp(4) & SO(11) & SO(10) & SU(4) & SU(5) \\
			\hline
			$Y_\chi$ & $1/3$ & $2/3$ & $2/3$ & $2/3$ & $2/3$ & $2/3$ \\
			$d_\chi$ & $16$ & $5$ & $11$ & $10$ & $6$ & $10$ \\
			\hline
	\end{tabular}
\end{center}
	\caption{Values of $d_\chi$ for the 12 models based on~\cite{Ferretti:2013kya} (see ~\cite{Belyaev:2016ftv,Cacciapaglia:2020vyf} for the  model list assignments). }
	\label{tab:M1_12}
\end{table}

The above results, however, were based on a spurion analysis where effective operators are constructed in terms of the spurions describing the partial compositeness mixing of the left-handed and right-handed tops. 
If the top partners are light compared to the other resonances, then the dominant contribution to the top mass may come from the direct mixing with the lightest baryon resonance. It was found in~\cite{Bizot:2018tds} that the two approaches based on operator analysis and on the baryon mixing are not completely equivalent: the couplings of a singlet pseudo-scalar were found to differ qualitatively and numerically. The main origin of this discrepancy traces back to  the fact that couplings to the light scalar only arise via the mixing terms, thus diagonalising the mass matrix is not equivalent to diagonalising the couplings.
Repeating the calculation in~\cite{Bizot:2018tds}  for the case of composite color octet couplings from the mixing with light top partners, the ratio $C_t/\kappa_g$ is obtained as
\beq
\frac{C_t}{\kappa_g} = \frac{n_{\chi L} \sin^2 \alpha_L + n_{\chi R} \sin^2 \alpha_R}{d_\chi}\,, 
\label{eq:ctokg2}
\eeq
where $\alpha_{L/R}$ are the mixing angles of the left- and right-handed tops to the composite states. 
Unlike in \eqref{eq:ctokg1}, where $n_{\chi L} = - n_{\chi R}$ yields full cancellation and thus a vanishing top coupling, \eqref{eq:ctokg2} implies that perfect cancellation becomes unlikely due to the dependence on the mixing angles, as typically $\sin \alpha_R > \sin \alpha_L$ to avoid bounds on the left-handed bottom. Thus, if the dominant contribution to the top mass may come from the direct mixing with the lightest baryon resonance, very small $C_t/\kappa_g$ is only achieved in models with $n_{\chi L} = n_{\chi R} = 0$ for $Y_\chi = 1/3$. 

This analysis shows that a small $C_t$ is not common in composite models, especially if the top mass comes from mixing to light top partners, but example models with naturally small $C_t$ exist. We also recall that the relation between top-coupling and the WZW term originates from a combination of demands: a composite Higgs, partial top compositeness and the color octet being the pNGB  of the color charged confined sector. The relation of the coupling to $gg$ and $\gamma g$ is more direct and only depends on the value of $Y_\chi$, as shown in Table~\ref{tab:octetBR}.

\section{Photons in color octet pair production: collider strategy} 

As we have seen in the previous section, photons from the color octet decays can play a very relevant role in the phenomenology, and searches in single production can be reinterpreted in this framework. However, no pair-production search based on photons exists so far. In this section we will cover this gap and establish a strategy to set up this kind of searches at the LHC (including the HL-LHC run) and at future higher-energy hadron colliders (FCC-hh).

In this pursuit, we are particularly interested in scenarios where the $\Phi\rightarrow gg$ is the dominant decay mode, thus we will set $C_t = 0$ ($\mbox{BR} (\Phi \to \bar{t}t) = 0$) in the following. While searches exist in the multi-jet final state for pair produced scalars, it must be noted that they are beset by a large irreducible QCD background. We will, therefore, consider final states with one or two photons and compare the sensitivity in these channels to the multi-jet one.
To facilitate the comparison, we define ratios of signal significance as follows
\begin{equation}
\delta_{ggg\gamma} \equiv \frac{Z_{ggg\gamma}}{Z_{gggg}} = \frac{S_{ggg\gamma}/\sqrt{B_{jjja}}}{S_{gggg}/\sqrt{B_{jjjj}}}\,, \quad 
\delta_{gg\gamma\gamma} \equiv \frac{Z_{gg\gamma\gamma}}{Z_{gggg}} = \frac{S_{gg\gamma\gamma}/\sqrt{B_{jjaa}}}{S_{gggg}/\sqrt{B_{jjjj}}}\,,
\label{eq:ratio}
\end{equation}
where the sensitivities $Z_x$ are simplistically defined as ratios of the number of signal events $S_x$ divided by the square root of the background events $B_x$. Note that we use $g$ and $\gamma$ to indicate parton level gluon and photon in the signal, while we use $j$ and $a$ to indicate detector reconstructed jet and photon in the background to take into account fakes.
We will use this ratio as an indicator of the relevance or dominance of the photon final states over the purely hadronic mode: it is evident in the simplistic definition of significance that the ratios are proportional to ratios if branching ratios. 
In the latter part of this section we will eventually adapt a more generalized version for the estimation of signal sensitivity, as we will discuss later. The analysis developed can be extended to any model that has a $g\gamma$ decay mode, thus we will give results for the benchmark models in Table~\ref{tab:octetBR}, as well as for general values of $BR(\Phi\rightarrow g\gamma) / BR(\Phi\rightarrow gg)$.

The first step in this direction corresponds to determining the backgrounds and the corresponding fake rates for the $jjja$ and $jjaa$ final states.  
The fake rate is due to the multi-jet background where a photon is radiated by the quark fragmentation or a jet is mistagged as a photon.  Given the large production cross sections for the multi-jet processes, these fake rates must be understood to a fair degree of accuracy. We begin with a detailed study of the different backgrounds relevant for our analysis.

\subsection{Background estimation} 

\begin{figure}[t]
	\begin{center}
			\includegraphics[width=7.2cm]{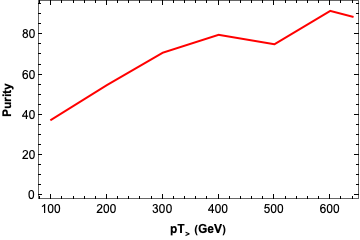}
	\end{center}
	\caption{Purity $N_{j\gamma}/(N_{j\gamma}+N_{jj})$ of the $ja$ samples with a single photon and a jet.}
	\label{fig:purity1}
\end{figure}

To correctly estimate the backgrounds, it is crucial to estimate the expected fake rates in the signal regions.
Since the multi-jet scenario is relatively complicated, as a proof of concept we first consider the purity of a $pp\rightarrow ja$ 
sample. To do so, we generated two samples of events: 
\begin{center}
1) $p p > j \gamma$ (pure sample); \; \; \; \; 2) $p p > j j$ (fake sample); \\
\end{center}
where $j=g,\ q$.
We simulate 150000 events at MonteCarlo level for both samples at $14$~TeV center of mass, requiring that the total $p_T$ of the two outgoing particles must be at least $300$~GeV. The events are generated using  {\tt{PYTHIA}} 8~\cite{Sjostrand:2007gs}, while we use {\tt{DELPHES}} 3~\cite{deFavereau:2013fsa} for the detector simulation. To include the effect of a jet mistagged as a photon, we use a flat probability $\epsilon = 10^{-4} $ in the {\tt{JetFakeParticle}} module of {\tt{DELPHES}}. Further, we assume the corresponding rate for the jets faking as either an electron or a muon to be zero.
The signal selection includes events with at least a single jet and a single isolated photon. Fig. \ref{fig:purity1} gives the purity fraction, defined as $N_{j\gamma}/(N_{j\gamma}+N_{jj})$, as a function of the lower $p_T$ threshold for the photon (denoted as $p_{T_>}$): all events with $p_T>p_{T_>}$ contribute and are normalized to the corresponding cross section. 
As expected, at lower thresholds we see a lower purity due to larger pollution from the $jj$ events. This is drastically reduced for increasing $p_{T_>}$ for the photon, while the upper value owes to the absence of high-$p_T$ photons in the samples. 
The purity for the $ja$ sample increases rapidly with the photon transverse momentum as the jet and the photon from the  $pp>j\gamma$ process is one QCD order lower than the one arising due to $pp>jj$.
As a validation, we can compare this behavior to the purity in the CMS analysis at $8$~TeV \cite{CMS:twa}, which shows good agreement. The absence of pile-up both in the data and in our simulations makes it a reasonable comparison.\footnote{The corresponding distribution is not available at $13$~TeV.}

\begin{figure}[t]
	\begin{center}
		\begin{tabular}{cc}
		\phantom{xxx} {$jjja$} & \phantom{xxx} {$jjaa$} \\
		\includegraphics[width=7.2cm]{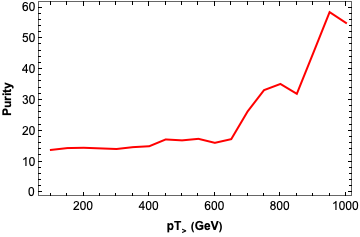}	&	\includegraphics[width=7.2cm]{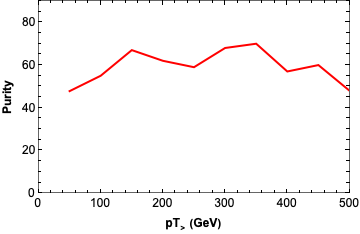}
		\end{tabular}
	\end{center}
	\caption{Purity of the $jjja$ (left) and $jjaa$ (right) samples. In the former, pollution comes from $jjjj$ events; for the latter, pollution derives from $jjj\gamma$ events.}
	\label{fig:purity2}
\end{figure}

Having validated our method, we now consider the purity of the $pp> jjja$ background. Events with a ``real'' photon primarily correspond to the radiation of a quark, therefore they are rather similar to events from the pure multi-jet sample.  As a result they are at the same QCD and QED order. Since we will consider the pair production of resonances of at-least $1$~TeV mass, we limit the matrix element generation of the background events to the following kinematical regions:
\begin{itemize}
\item Multi-jet: The  scalar $p_T$ sum of the outgoing partons is required to be at least $1500$~GeV;
\item $pp> jjja$ The scalar $p_T$ sum of the outgoing partons is required to be at least $1400$~GeV, while the photon is required to have a $p_T$ of at-least $100$~GeV.
\end{itemize}
As emphasised earlier, since the $jjjj$ and the $jjj\gamma$ samples are characterized by Feynman diagrams at the same order, the purity in this instance is mainly a measure of the percentage of the jet faking a photon as well as possibility of a hard radiated photon being isolated from the multi-jet sample. With increasing photon $p_T$, the probability of a hard radiated photon from the $jjjj$ sample decreases.
The left plot of Fig.~\ref{fig:purity2} illustrates the purity for the $jjja$ sample in our simulation.
 We accept events with at-least three jets and an isolated photon from both multi-jet and $jjj\gamma$ samples.
 Owing to the similarity between the two, the multi-jet sample dominates for low and intermediate $p_T$ on account of its greater cross-section, thus the multi-jet background will dominate for low mass resonances. With increasing $p_T$, one observes an increase in the purity due to the fact that the photon from the $jjj\gamma$ samples has a mildly larger tail than the one due to $jjjj$ (the latter sensitive to the ``softer'' Bremsstrahlung). 
The fluctuations at the tail of the curve can be attributed to a lack of statistics. 
Our analysis shows, therefore, that the multi-jet background is relevant for the $ggg\gamma$ signal and the results of our later analysis must be taken with caution: in fact, only data driven techniques allow for a reliable estimate of this background, while our method is limited by statistics and the accuracy of the MonteCarlo in characterizing the tails of the event distributions.

Finally, to account for the possibility of the color octet pair decaying in two photons and two gluons, we estimate the corresponding $jjaa$ background. In this case, the ``pure'' background is due to $p p > jj\gamma\gamma$, with contamination from the $p p > jjj\gamma$ processes. The background events $jj\gamma\gamma$ are similar to the di-jet events considered earlier with the additional radiation of photons off the quarks. To match the kinematics of the signal, we further require that the photons have  $p_T >150$~GeV at particle level, while the scalar $p_T$ sum of the outgoing partons is required to be at-least $500$~GeV. The fakes are mainly due to $jjj\gamma$ events, with a jet faking a photon or hard radiation in a jet.
The pure multi-jet no longer contributes to this final state, where we estimated a contamination probability of $ <10^{-5}$. Given the small cross section for $p p > jj\gamma\gamma$, it is important to carefully estimate the contributions from $p p > jjj\gamma$. 
The right plot of Fig.~\ref{fig:purity2} gives the purity fraction for the $jjaa$ background from our simulation
as a function of the $p_T$ of the subleading photon. Similarly to $ja$ purity in Fig.~\ref{fig:purity1}, it exhibits a plateau at intermediate $p_T$ regime. However, it also exhibits a gradual decline with increasing $p_T$: this can be attributed to the fact that the photons in $p p > jj\gamma\gamma$ events are mainly due to radiation and are not necessarily characterised by high $p_T$, while a high $p_T$ jet faking a photon is more probable for $p p > jjj\gamma$. This implies that both backgrounds have to be considered for the signal corresponding to the $jjaa$ final state.

\subsection{Signal and background acceptances}

\begin{table}[t]
	\begin{center}
		\small\begin{tabular}{|l|c|c|c|} 
			\hline
 & $jjjj$  & $jjja$ & $jjaa$\\ \hline
			\hline
\phantom{\Big(} final state & $N_j \geq 4$ & $N_j \geq 3,\;\; N_a = 1$ & $N_j \geq 2,\;\; N_a = 2$ \\ \hline
\phantom{\Big(} $p_T$ cuts & $p_T^j > 80$~GeV & \multicolumn{2}{c|}{$p_T^j > 150$~GeV, \;\; $p_T^a > 50$~GeV} \\ \hline
\phantom{\Big(} pairs id. & \multicolumn{2}{c|}{min. $\sum_{i=\text{pairs}} |\Delta R_{i}-0.8|$ } & min. $\Delta R_{a_1 j_i}$\\ \hline
\hline
\phantom{\Big(}  asymmetry &   $\frac{|m_{jj_1} - m_{jj_2}|}{m_{jj_1} + m_{jj_2}} < 0.1$ & $\frac{|m_{ja} - m_{jj}|}{m_{ja} + m_{jj}} < 0.1$ & \\ 
\phantom{\Big(} parameters &   $|\eta_{jj_1} - \eta_{jj_2}| < 0.1$  & $|\eta_{ja} - \eta_{jj}| < 0.1$ & \\ \hline 
\hline
\phantom{\Big(} binning &  none & $m_{ja}$ & $m_{ja_1}$ \\ \hline
		\end{tabular} \normalsize
	\end{center}
	\caption{ Cut flow for the 3 final states, where the multi-jet one follows closely the CMS pair di-jet search of \cite{Sirunyan:2018rlj}.}
	\label{tab:cutflow}
\end{table}

With an understanding of the different backgrounds, we proceed to estimating background and signal efficiencies corresponding to the event selection requirements.
For the signal, we consider the following benchmark points, expressed in terms of masses and $14$~TeV pair production cross section, $(M_\Phi\ [\mbox{GeV}], \sigma_{\rm pair}\ [\mbox{fb}])$: 
\begin{eqnarray}
\text{BP1}::(900,74.2),\;\; \text{BP2}::(1000,34.4),\;\;\text{BP3}::(1100,16.7),\;\;\text{BP4}::(1200,8.3).\nonumber 
\label{eq:bp}
\end{eqnarray}
These points are not yet excluded by the current searches, as shown in the previous section, in particular by the pair-dijet resonance search. To validate our analysis, we also consider a point at the edge of the excluded mass range:
\begin{eqnarray}
\text{BP0}::(700, 400). \nonumber
\end{eqnarray}
Like for the background, the parton level signal events are simulated at $14$~TeV using MADGRAPH~\cite{Alwall:2014hca} and showered by PYTHIA 8~\cite{Sjostrand:2007gs}. We use the CMS card for DELPHES 3~\cite{deFavereau:2013fsa} for the detector simulation.
The jets are reconstructed using FASTJET~\cite{Alwall:2014hca}, following the anti-kt algorithm with $R=0.4$ and $p_T=20$~GeV.
 
First, for the multi-jet final state, we closely follow the CMS pair-dijet search of \cite{Sirunyan:2018rlj} by means of the cuts outlined in the second column of Table~\ref{tab:cutflow}. We select events with at least 4 jets with $p_T > 80$~GeV.  In order to select the two best di-jet pairs compatible with the signal, the four leading jets, ordered in $p_T$, are combined to create three unique combinations of di-jet pairs per event. Out of the three combinations, the di-jet configuration with the smallest $\Delta R_{dijet}=\sum_{i=1,2}|\Delta R_i-0.8|$ is chosen, where $\Delta R^i$ is the distance in the $\eta-\phi$ plane between the two jets in the $i^{th}$ di-jet pair.
Once the best pairing is selected, two asymmetry parameters are defined, as in Table~\ref{tab:cutflow}, to reduce the QCD background, where $m_{jj_i}$ and $\eta_{jj_i}$ are the invariant mass and total pseudo-rapidity of the $i^{th}$ jet pair.
To provide realistic estimates of the sensitivities, as the multi-jet background is hardly modelled by MonteCarlo generators, we decided to use instead the data-driven estimates used in the CMS search, shown in Fig. 9 of \cite{Sirunyan:2018rlj} for an accumulated luminosity of $35~fb^{-1}$.

For the final state $jjja$, given the absence of resonance searches, we adopt a methodology similar to the multi-jet searches: events are selected with one isolated photon with $p_T > 50$~GeV and at least 3 jets with $p_T > 150$~GeV. The best pairing of the photon with a jet is selected by use of the same technique as above, and furthermore we employ the $jjjj$ asymmetry variables with one $jj$ pair replaced by the $ja$ one. This approach was particularly useful in limiting the multi-jet QCD background, which we found to be significant in the $jjja$ channel as shown in the purity plot of Fig.~\ref{fig:purity2}.
The cut-flow is summarized in the third column of Table~\ref{tab:cutflow}. Finally, in order to extract the sensitivities, we bin the events in the $m_{ja}$ distribution following the best pairing method described above.
In Table~\ref{tab:eff619} we show the acceptance on signal and background after the cuts based on the asymmetry parameters, defined as the ratio of events that pass the cuts over the total number of generated events.  In this instance the number for the both the signal and background indicates events which pass the selection due to the asymmetry parameters.
We see that the signal acceptance for the multi-jet and $jjja$ final states are similar.

\begin{table}[t]
	\begin{center}
		\small\begin{tabular}{|c|c|c|c|c|} 
			\hline
			
			&$jjjj$  & $jjja$ & \multicolumn{2}{c|}{$jjaa$}\\ \hline
			\hline
			
			Signal acceptance $\epsilon_S$ ($M_\Phi=1$~TeV)        &$0.008$&$0.008$& \multicolumn{2}{c|}{$0.67$}\\ \hline   
			Background acceptance $\epsilon_B$&Data&$0.002$&$0.69$ & $0.009$\\\hline
			Background cross section (fb)&Data&$1.34\times 10^{3}$&$7$ &  $1.34\times 10^{3}$\\\hline
		\end{tabular} \normalsize
	\end{center}
	\caption{ Acceptances for signal and backgrounds for the three final states we study, after the cuts in Table~\ref{tab:cutflow}.
	The signal corresponds to BP2, but the acceptances are rather independent on the mass value, while the background cross sections ae
	calculated after imposing a cut on the scalar sum of the $p_T$ of the outgoing particles $p_{T,\text{sum}} > 1500$~GeV.
	For $jjaa$, we show separately for the physical background $jj\gamma\gamma$ and the one from $jjj\gamma$ via jet mistagging.}
	\label{tab:eff619}
\end{table}

\begin{figure}[t]
	\centering
\begin{tabular}{cc}
	\phantom{xxx} $M_\Phi = 900$~GeV & \phantom{xxx} $M_\Phi = 1200$~GeV \\
	\includegraphics[width=7cm,height=7cm,keepaspectratio]{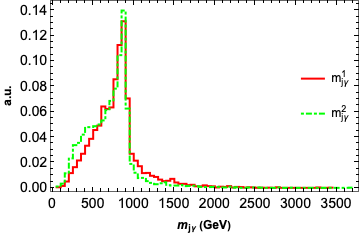} &
	\includegraphics[width=7cm,height=7cm,keepaspectratio]{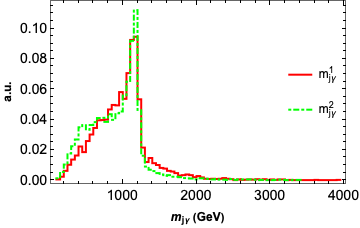}
\end{tabular}
	\caption{Distribution of the invariant masses $m_{j\gamma}$ of the two pairs for $M_{\Phi}=900$~GeV (left) and $1200$~ GeV (right). See text for description.}
	\protect\label{fig:pairinvmass}
\end{figure}

The final state $jjaa$ is relatively simpler as there are only two combinatorial possibilities for the invariant mass reconstruction. Furthermore, the requirement of two isolated photons is sufficient to limit the multi-jet background to an insignificant amount. This makes it possible to adopt simpler selection criteria than the ones defined for multi-jet final state, and it is possible to increase the acceptance for signal events.  
We thus select events with at least two jets and exactly two isolated photons: this selection alone leads to the acceptances listed in the last column of Table~\ref{tab:eff619}. For the backgrounds, we listed separately the one coming from events with two real photons, and the one from $jjj\gamma$: the latter is highly suppressed due to the low fake rate.
Because of the natural suppression of the QCD background, there is no need to impose the cuts from the asymmetry parameters.\footnote{In principle, it would be possible to increase the significance by adding cuts on the asymmetry parameters, however this would reduce too much the background events in our simulation.}
We, therefore, content ourselves with identifying the correct pairing of photons and jets: after ordering the photons in $p_T$, we calculate the angular distance from the two jets and select the jet with the minimal value.
The effectiveness of this strategy is shown in Figure \ref{fig:pairinvmass}, where we show the invariant masses of the two pairs for BP1 ($M_\Phi = 900$~GeV) and BP4 ($M_\Phi = 1200$~GeV). As it can be seen, both distributions nicely peak on the physical mass of $\Phi$.

\subsection{Signal sensitivities}

With an estimation of the collider efficiencies for the signal and the background, we are in a position to compute the respective signal sensitivity. It must be pointed here that the acceptances for signal and backgrounds in Table \ref{tab:eff619} are calculated naively using  the 
events which satisfy the corresponding signal selection criteria. Since the mass of the underlying resonance is an unknown parameter, it is beneficial to define a variable sensitive to local variations in event multiplicities without biasing oneself to 
restricted regions of signal phase space.
This can be put into practice by the  definition of the following binned sensitivity \cite{Cowan:2010js}:
\begin{equation}
Z^{\rm bin}=\sqrt{\sum_i \left(2(s_i+b_i)\log\left[1+\frac{s_i}{b_i}\right]-2s_i\right)}\,,
\label{eq:sensitivity}
\end{equation}
where $i$ runs over the bins and computes the signal and background events for a given observable in each bin. This facilitates a comparison of the signal and background events in each bin and is sensitive to the presence of signal events reconstructed away from the pole mass. This takes into account signal events which could have ordinarily been missed due to mass selection around the pole. Moreover it offers a fairly democratic search strategy as the mass of the underlying resonance is unknown.
 Note that, since the variable $Z^{\rm bin} $ is a bin wise comparison, the bins which do not contain signal events do not contribute to the sensitivity and hence do not affect its computation.

\begin{table}[t]
	\begin{center}
		\small\begin{tabular}{|c|c|c|c|} 
			\hline
			
			&4$j$  & 3$j$+1$\gamma$& 2$j$+ 2$\gamma$\\ \hline
			\hline
			
			 $Z$         &$2.2$&$1.1\ (3.23)$&$8.49\ (9.84)$\\ \hline   
\end{tabular} \normalsize
	\end{center}
	\caption{ Comparison of signal sensitivities for the BP0 between different channels. The numbers are quoted for $35~\mbox{fb}^{-1}$ of integrated luminosity. The number in bracket for the photon mode corresponds to bin-wise estimation of the signal sensitivity. See text for descriptions}
	\label{tab:eff3619}
\end{table}

To compare the sensitivity of the final states with photons to the multi-jet one, we make use of the ratios defined in Eq.~\eqref{eq:ratio}, where the sensitivity in the $jjja$ and $jjaa$ channels are computed using the binned formula of Eq.~\eqref{eq:sensitivity}. For the multi-jet final state, $jjjj$, since we use the CMS data at $35~\mbox{fb}^{-1}$, we simply use $Z_{gggg} = S/\sqrt{B}$ as the background is not known for different bin sizes.
To validate our method, we fist focus on the benchmark point BP0, with branching ratios corresponding to the benchmark model with $Y_\chi = 2/3$ (see Table~\ref{tab:octetBR}), for which $\mbox{BR} (\Phi \to gg) \approx 80\%$ and $\mbox{BR} (\Phi \to g\gamma) \approx 15\%$.
The signal sensitivities for the different channels are given in Table~\ref{tab:eff3619} for an integrated luminosity of $35~\mbox{fb}^{-1}$, corresponding to the current CMS di-jet pair search.
We first calculate the sensitivities using $Z = S/\sqrt{B}$, counting events in a mass window around the resonance, in a similar way for the three final states. Then, we indicate with numbers in brackets the results obtained with the binned sensitivity of Eq.~\eqref{eq:sensitivity}, where the bin size is $100$~GeV, and the distributions correspond to $m_{j\gamma}$ for $jjja$ and $m_{j\gamma_1}$ for $jjaa$ (invariant mass of the pair containing the highest $p_T$ photon). We also tested the stability against the bin size by varying it between $80$ and $150$~GeV without noticing any significant variation.\footnote{Smaller bin sizes were not allowed for lack of statistics in the background events of our simulation.} 
Our results show that, even without the binned sensitivity, the final state with two photons offers a much stronger reach compared to the multi-jet final state. Furthermore, the binning gives a significant increase in sensitivity, and we will use it in the following estimates.

\begin{table}[t]
	\begin{center}
		\begin{tabular}{|ccc|ccc|} 
			\hline
				\multicolumn{3}{| c| }{BP1\; (900,74.2)}&\multicolumn{3}{ c| }{BP3\; (1100,16.7)}\\
			& $Y_\chi=1/3$  & $Y_\chi=2/3$&           & $Y_\chi=1/3$  & $Y_\chi=2/3$  \\ \hline

			$Z_{gggg}$         &\begin{tabular}{c}10.32 \\\end{tabular}&\begin{tabular}{c}7.48 \\\end{tabular}&$Z_{gggg}$&\begin{tabular}{c}4.8 \\\end{tabular}&\begin{tabular}{c}3.54 \\\end{tabular}\\ \hline

			$Z_{ggg\gamma}$         &$5.02$&$13.03$&$Z_{ggg\gamma}$&$1.09$&$3.06$\\    
			$\delta_{ggg\gamma}$ & $\bf 0.48$ & $\bf 1.74$ & $\delta_{ggg\gamma}$&$\bf 0.22$ & $\bf 0.86$ \\ \hline
			
			$Z_{gg\gamma\gamma}$     &$1.91$&$21.08$&$Z_{gg\gamma\gamma}$&$0.47$&$5.31$ \\ 
			$\delta_{gg\gamma\gamma}$ & $\bf 0.18$&$\bf 2.81$ & $\delta_{gg\gamma\gamma}$ & $\bf 0.09$ & $\bf 1.5$ \\ \hline  \hline

			\multicolumn{3}{| c| }{BP2\; (1000,34.4)}&\multicolumn{3}{ c| }{BP4\; (1200,8.3)}\\
			& $Y_\chi=1/3$  & $Y_\chi=2/3$&           & $Y_\chi=1/3$  & $Y_\chi=2/3$  \\ \hline

		  $Z_{gggg}$         &\begin{tabular}{c}7.5 \\\end{tabular}&\begin{tabular}{c}5.45 \\\end{tabular}&$Z_{gggg}$&\begin{tabular}{c}2.6 \\\end{tabular}&\begin{tabular}{c}1.91 \\\end{tabular}\\  \hline

			$Z_{ggg\gamma}$         &$2.19$&$6.07$&$Z_{ggg\gamma}$&$0.63$&$1.77$\\   
			$\delta_{ggg\gamma}$ & $\bf 0.29$& $\bf 1.11$ &$\delta_{ggg\gamma}$&$\bf 0.24$&$\bf 0.92$ \\ \hline

			$Z_{gg\gamma\gamma}$     &$0.98$&$10.97$&$Z_{gg\gamma\gamma}$&$0.25$&$2.83$\\ 
			$\delta_{gg\gamma\gamma}$ & $\bf 0.13$ & $\bf 2.01$ &$\delta_{gg\gamma\gamma}$ & $\bf 0.09$ & $\bf 1.48$ \\ \hline 
		\end{tabular}
		\caption{Estimation of signal sensitivities ($Z$) and ratio $\delta$ for different channels at HL-LHC with $3~\mbox{ab}^{-1}$ of integrated luminosity. The multi-jet background at this luminosity is  obtained by scaling the events multiplicities from $35~\mbox{fb}^{-1}$. The benchmark points are denoted by $(M_\Phi~\text{[GeV]},\sigma_{prod}~\text{[fb]})$. } 	\label{tab:eff2619}
	\end{center}
\end{table}

We now turn our attention to the benchmark points BP1-BP4, with masses extended beyond the current exclusion. 
We first computed the expected sensitivities at the HL-LHC for an integrated luminosity of $3~\mbox{ab}^{-1}$ for the benchmark models in Table \ref{tab:octetBR}.
For the multi-jet final state, we estimate the background by rescaling the ones from the CMS analysis at $35~\mbox{fb}^{-1}$ to the higher luminosity.
The results are shown in Table~\ref{tab:eff2619}, where the bold numbers correspond to the ratios defined in Eq.~\eqref{eq:ratio}.
The values clearly show that, for the pessimistic case $Y_\chi = 1/3$, the multi-jet final state always gives stronger reach having $\mbox{BR} (\Phi \to gg) \approx 90\%$. Instead, the optimistic case shows that it is the $jjaa$ final state that provides the best sensitivity. In both cases, sensitivities close to $3\sigma$ exclusion can be obtained, so that color octet masses up to $1.2$~TeV seem to be reachable at the HL-LHC (note that a combination of the three final states can further improve the reach).

\begin{figure}[t]
	\begin{center}
		\includegraphics[width=7.2cm]{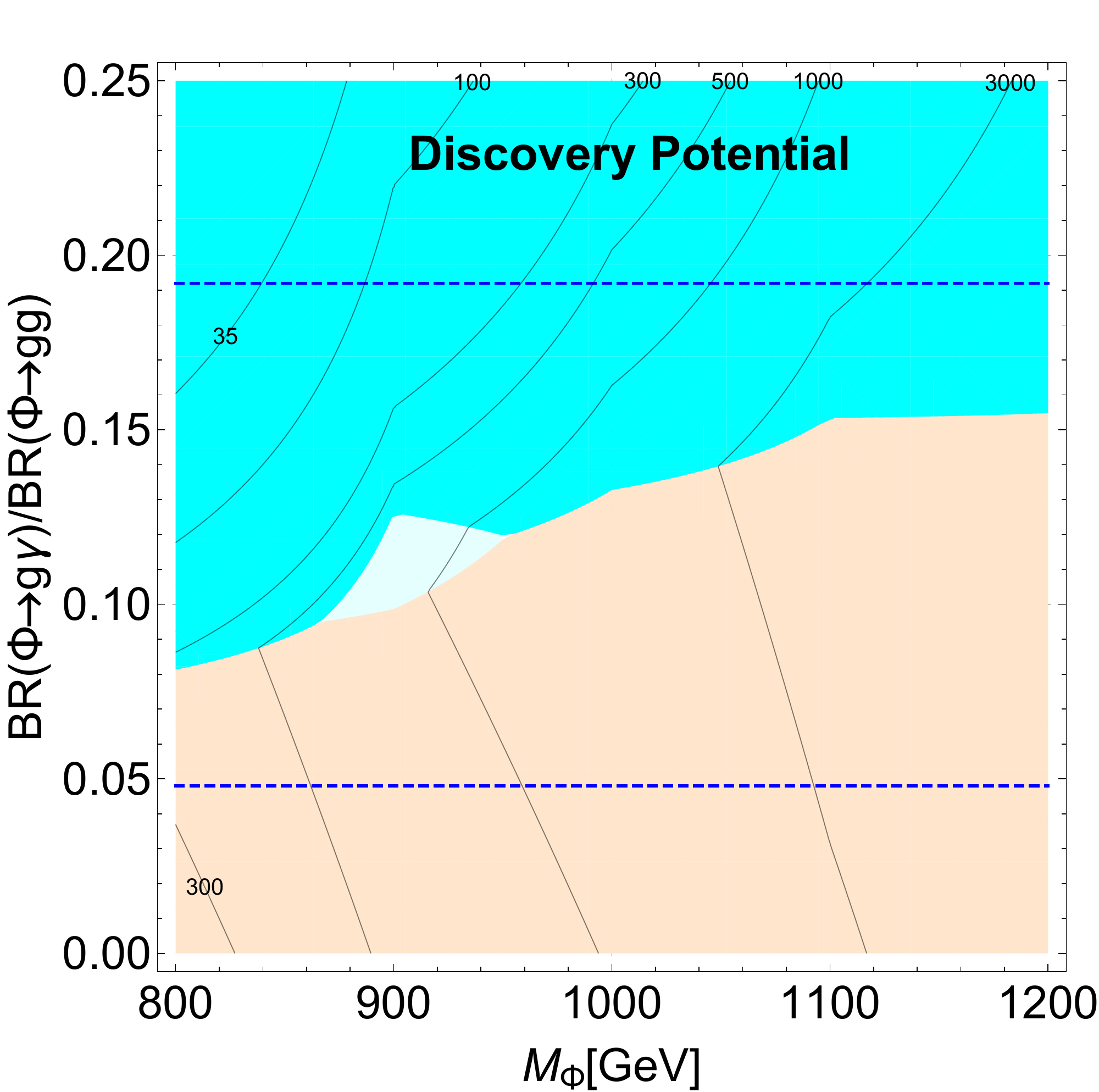}	\includegraphics[width=7.2cm]{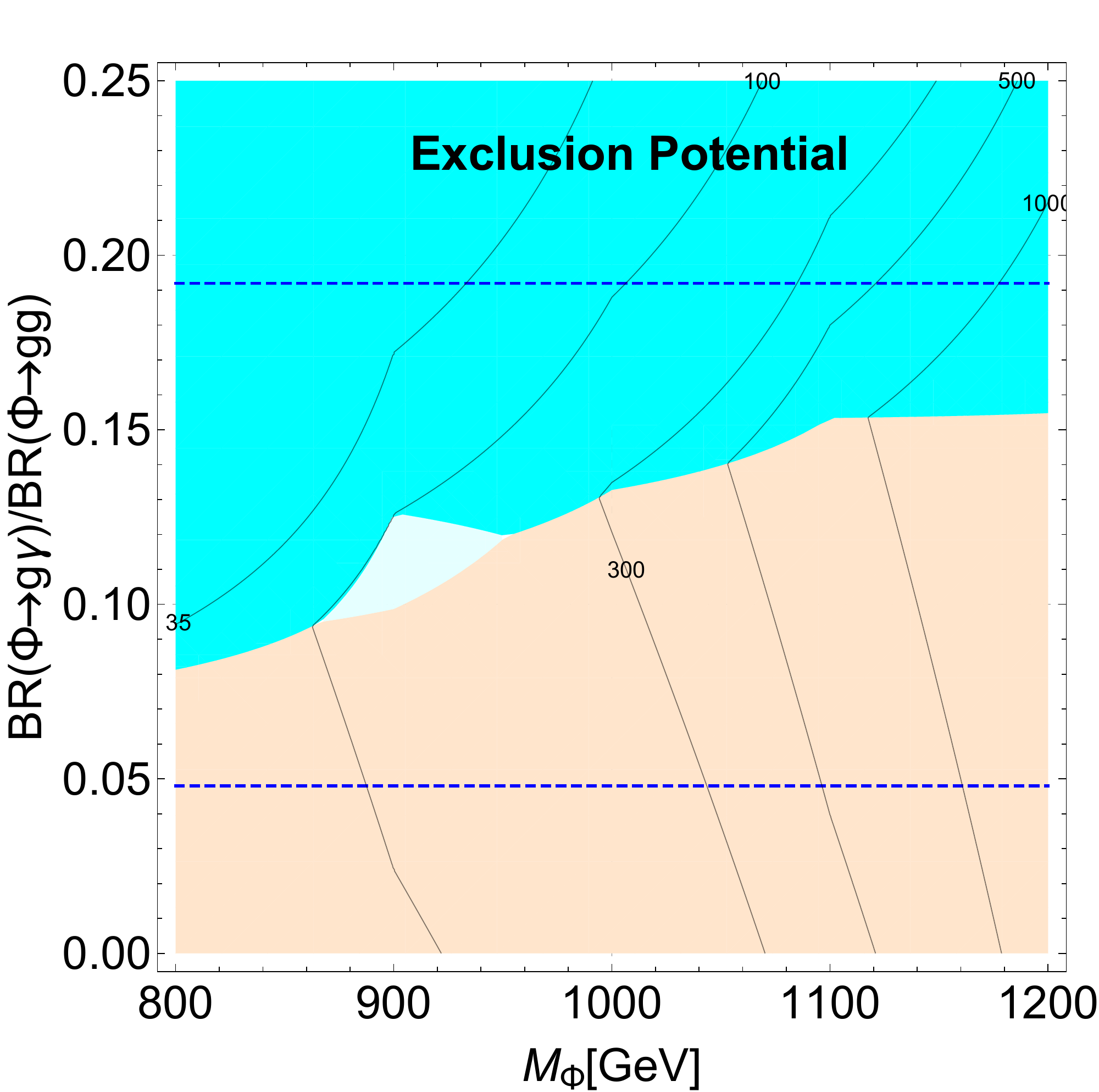}
	\end{center}
	\caption{ Contours of the luminosity with expected $Z = 5$ (``Discovery Potential'') on the left and $Z = 2$  (``Exclusion Potential'') on the right, as a function of $M_\Phi$ and the relative branching fraction $BR(\Phi\rightarrow g\gamma) / BR(\Phi\rightarrow gg)$, based on the HL-LHC analysis strategy  in this subsection. The most sensitive channels are expected to be $gg\gamma\gamma$ (dark cyan), $ggg\gamma$ (light cyan), or $gggg$ (orange).}
	\label{fig:discexcl}
\end{figure}

The pattern of the ratios $\delta$ suggests that the relative dominance of a given mode depends on the relative branching fraction as well as the mass of the underlying resonance. This aspect is illustrated in Fig.\ref{fig:discexcl} where we extend the  analysis  to  generic  ratios $BR(\Phi\rightarrow g\gamma) / BR(\Phi\rightarrow gg)$. The two plots show contours of the integrated luminosity corresponding, on the left, to the ``discovery potential'' ($Z = 5$) and, on the right, to the ``exclusion potential'' ($Z=2$), as a function of $M_\Phi$ and the relative branching fraction.  In the dark cyan areas, the $gg\gamma\gamma$ channel has best projected sensitivity, while in the orange areas $gggg$ is expected to dominate. The (small) light cyan area indicates best sensitivity of $ggg\gamma$ (which in this area is slightly stronger, but comparable to the other channels). The lighter masses exhibit a better reach of the $gg\gamma\gamma$ channel even for smaller values of the ratio $BR(\Phi\rightarrow g\gamma) / BR(\Phi\rightarrow gg)$. This can be attributed to the significant multi-jet background in the lower invariant mass bins. However, with increasing $M_\phi$ the background rate drops more rapidly than for the $gg\gamma\gamma$ channel (for which less severe cuts have been applied). The projection for heavier masses is limited on two accounts: lack of background statistics and rapidly dropping signal cross sections. The advent of future proton colliders would serve as a natural continuation to probe these states and will be discussed in the next section.

\subsection{The next collider: FCC-hh}

\begin{figure}[t]
	\begin{center}
		\includegraphics[width=7.2cm]{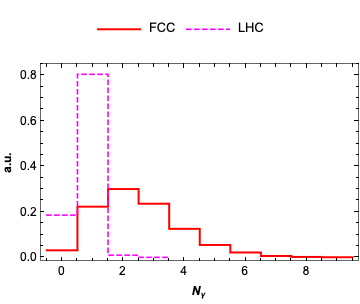}
	\end{center}
	\caption{Comparison of photon multiplicities $N_\gamma$ for the $ggga$ signal events between the HL-LHC and the FCC-hh. }
	\label{fig:photondist}
\end{figure}

The analysis in the previous sections shows that the HL-LHC is expected to have good sensitivity to probe masses $M_\Phi \leq 1.2$~TeV, and that adding searches for final states with one and two photons can help in significantly extending the reach. 
This analysis offers a natural progression leading to the exploration of higher masses at a future hadronic collider, like the FCC-hh at a center-of-mass energy of $100$~TeV. While the FCC can be expected to probe masses as heavy as tens of TeV, as a first test we estimated the reach for the following two benchmark points, $(M_\Phi\ [\mbox{TeV}], \sigma_{\rm pair}\ [\mbox{fb}])$: 
\begin{eqnarray}
\text{BP5}::(2.5,270),\;\; \text{BP6}::(3,100)\,;\nonumber 
\label{eq:bp}
\end{eqnarray}
where the cross sections correspond to the FCC-hh energies.
We simulated both signal and backgrounds using the standard FCC-hh cards, where
the background samples in this phase space is generated by requiring the scalar $p_T$ sum of the outgoing particles to be at-least $2$~TeV.
One key difference between the analyses at the two machines is the difference in the photon multiplicities. For the HL-LHC, the event selection was associated with the requirement of exactly one or two isolated photons, satisfying a certain minimum $p_T$ criterion that corresponds to the signal. However, given the larger center of mass energy at the FCC, the radiated photons are also expected to be both isolated and hard. The difference in photon multiplicities, after the cuts, between the two machines is illustrated in Fig. \ref{fig:photondist}. As a result, the requirement of exactly one or two isolated photons would diminish the signal acceptance. In light of this the event selection is slightly modified with respect to the HL-LHC one, with the requirement of at-least a single isolated photon. However, the signal acceptance efficiencies after the cuts is machine independent and roughly mirrors the values given in Table \ref{tab:eff619}. 

\begin{figure}[t]
	\begin{center}
		\includegraphics[width=7.2cm]{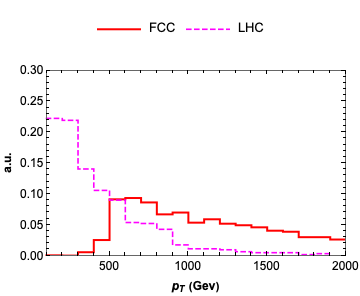}	\includegraphics[width=7.2cm]{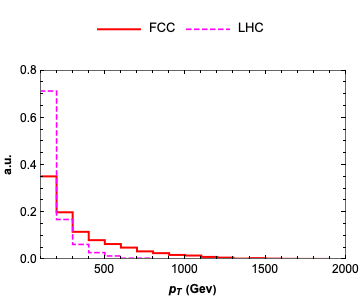}
	\end{center}
	\caption{Comparison of $p_T$ distribution for the $jjja$ background between the HL-LHC and the FCC-hh. The left plot is for the leading $p_T$ photon and the right plot is for the sub-leading one. }
	\label{fig:photonpTdist}
\end{figure}

\begin{table}[b]
	\begin{center}
		\begin{tabular}{|ccc|ccc|} 
			\hline
			\multicolumn{3}{| c| }{BP1\; (2.5,270)}&\multicolumn{3}{ c| }{BP2\; (3,100)}\\
			& $Y_\chi=1/3$  & $Y_\chi=2/3$&           & $Y_\chi=1/3$  & $Y_\chi=2/3$  \\ \hline
			
			$Z_{ggg\gamma}$         &$11.7$&$30.8$&$Z_{ggg\gamma}$&$4.98$&$13.6$\\ \hline

			$Z_{gg\gamma\gamma}$     &$5.47$&$58.2$&$Z_{gg\gamma\gamma}$&$2.87$&$31.1$\\ \hline 
		\end{tabular}
		\caption{Estimation of signal sensitivities ($Z$) for different channels at $3~\mbox{ab}^{-1}$ of integrated luminosity for the FCC. We recall that the numbers for the $jjj\gamma$ signal is underestimated as the mulit-jet background has not been considered in its evaluation.  } \label{tab:effFCC}
	\end{center}	
\end{table}

There is also another key difference between the analyses at the two machines: the photon $p_T$ spectrum for the background. Since there is a larger probability for the radiated photons to pass the isolation criteria at the FCC, it is also instructive to compare its $p_T$ distribution with respect to the HL-LHC. For illustration we consider the $jjj\gamma$ background, and compare the $p_T$ of the leading and the sub-leading $p_T$ photons at the two colliders. Fig. \ref{fig:photonpTdist} illustrates this difference for the leading photon (left) and the sub-leading photon (right). Note the significantly longer tail at the FCC, which eventually leads to the presence of background in larger invariant mass bins. This suggests that for the $jjaa$, the $jjj\gamma$ background is more relevant at the FCC. The same argument can be also be extended to the relevance of the multi-jet background for the signal with one photon and three jets. Given the computational limitations, we cannot accurately represent the multi-jet backgrounds, thus we do not consider them in our estimates for computing the signal acceptance. As a result the numbers for the $jjja$ final state are likely to be overestimated. Nevertheless, since the multi-jet background is unlikely to generate two hard photons, our estimates for the  $jjaa$ final state ($p^\gamma_T> 150$ GeV) are likely to be accurate. The corresponding values of sensitivities are given in Table \ref{tab:effFCC},  we use mass bins of width $350$~GeV to avoid fluctuations in background statistics.
Our results clearly show that the FCC-hh will have an excellent reach for a color octet by looking at final states with photons.

\section{Conclusion}

Heavy colored composite states are a feature of several frameworks beyond the SM. A  coupling to the gluons, proportional to $\alpha_s$, ensures that their decay into the $gg$ final state is the most dominant mode. This would naturally motivate searches for these states in the multi-jet final state. In this work we invoke the possibility of the coupling of these states to a gluon and a photon through the WZW term.  In the pair production of these states, we show that final states with two photons (and to a minor extent with one photon) could be more sensitive than the multi-jet final state. Depending on the underlying mass and the ratio $BR(\Phi\rightarrow g\gamma) / BR(\Phi\rightarrow gg)$, we identify the most dominant mode for their discovery (or for setting stronger exclusions).  The reach for the HL-LHC extends up to $1.2$~TeV. Masses beyond this range are better accessed at a FCC, and we provide preliminary estimates for a center-of-mass energy of $100$~TeV. This study does not only strongly motivate the  exploration of pair produced color octets with one and two photons, but also does offer a natural progression from the LHC to future proton colliders.

 \acknowledgments

 We wish to thank V. Sordini for useful comments and discussions, and comments on the manuscript.
 AI, GC, AD would like to thank CEFIPRA Indo-French research network.
GC and AD acknowledge partial support from the France-Korea Particle Physics Lab (FKPPL) and the Labex-LIO (Lyon Institute of Origins) under grant ANR-10-LABX-66 (Agence Nationale pour la Recherche), and FRAMA (FR3127, F\'ed\'eration de Recherche ``Andr\'e Marie Amp\`ere'').
TF's work is supported by IBS under the project code IBS-R018-D1.
GC, AD and TF would like to express a special thanks to the Mainz Institute for Theoretical Physics (MITP) of the Cluster of Excellence PRISMA+ (Project ID 39083149) for its hospitality and support during part of this work.

\appendix

\section{Bookkeeping of neutral color octet models} \label{app:bookkeeping}

In this Appendix we present a classification of models that contain a color octet pseudo-scalar. The key ingredient is to add information about the electroweak quantum numbers of the multiplet the neutral state belongs to. We will classify the cases in terms of the weak isospin $I$, $\Phi_{I}$, with $I = 1, 2, 3$ as minimal cases, and focus on the neutral pseudo-scalar components matching the effective Lagrangian in Eq.~\eqref{eq:Lag2}.

\subsubsection*{Isospin 1}

In this case, the multiplet has a single neutral component, $\Phi_1$. The leading order couplings to tops and gauge bosons arise at dim-5 level, and can be written in a gauge-invariant way as
\begin{equation}
\mathcal{L}_{\Phi_1} =  i\ \Phi^a_1\ \left(\frac{\lambda_t}{f_{\Phi}} \varphi_{H}^\dagger \bar{q}_L \frac{\lambda^a}{2} t_R - \mbox{h.c.} \right)
+ \frac{\alpha_s \kappa_{g}}{8 \pi f_{\Phi}}\Phi^a_1\ \epsilon^{\mu\nu\rho\sigma}\left[ \frac{1}{2}d^{abc}\ G^b_{\mu\nu}G^c_{\rho\sigma}+\frac{{g'} \kappa_B}{{g_s}  \kappa_{g}}\ G^a_{\mu\nu}B_{\rho\sigma}\right],
\end{equation}
where $\varphi_H$ is the Higgs doublet and SU(2)$_L$ contractions are left understood.
After the Higgs field develops a vacuum expectation value $\langle \varphi_H^\dagger \varphi_H \rangle = v^2/2$, we have the following effective couplings
\begin{equation}
C_t = \frac{\lambda_t v}{\sqrt{2} m_t}\,, \quad \kappa_\gamma = \kappa_Z = \kappa_B\,.
\end{equation}
This case occurs in models of top partial compositeness with an underlying gauge-fermion description~\cite{Cacciapaglia:2015eqa,Belyaev:2016ftv}.

\subsubsection*{Isospin 2}

In this case, the neutral color octet belongs to a doublet with hypercharge $Y_\Phi = 1/2$, which also contains a charged component and a neutral scalar:
\begin{equation}
\Phi_2 = \begin{pmatrix} \Phi_2^+ \\ (\tilde{\Phi}_{2,0} + i \Phi_{2,0} )/\sqrt{2} \end{pmatrix}\,.
\end{equation}
This case is interesting as a scalar extension of the SM~\cite{Manohar:2006ga}, as it does not suffer from large flavour changing neutral currents~\cite{Arnold:2009ay}.
Now, it is possible to write a dim-4 coupling to tops:
\begin{equation}
\mathcal{L}_{\Phi_2} \supset \lambda_t (\Phi_2^\dagger)^a\ \bar{q}_L \frac{\lambda^a}{2} t_R + \mbox{h.c.}
\end{equation}
which gives a large $C_t = \frac{\lambda_t f_\Phi}{\sqrt{2} m_t}$, plus similar couplings of the real and charged components. Thus, if the coupling to tops is present, decays will be dominated by $t\bar{t}$.

Couplings to gauge bosons can only arise at dim-6, in the form:
\bea
\mathcal{L}_{\Phi_2} &\supset& \phantom{+}\frac{g_s^2 c_g}{32 \pi^2 f_\Phi^2} \left( i \varphi_H^\dagger \Phi_2^a + \mbox{h.c.} \right)\ \epsilon^{\mu\nu\rho\sigma}  \frac{1}{2}d^{abc}\ G^b_{\mu\nu}G^c_{\rho\sigma}  \nonumber \\
&&+ \frac{g_s {g'} c_B}{32 \pi^2 f_\Phi^2}   \left( i \varphi_H^\dagger \Phi_2^a + \mbox{h.c.} \right) \ \epsilon^{\mu\nu\rho\sigma}  G^a_{\mu \nu} B_{\rho \sigma}  \nonumber \\
&&+ \frac{g_s {g} c_W}{32 \pi^2 f_\Phi^2}   \left( i \varphi_H^\dagger \tau^\alpha \Phi_2^a + \mbox{h.c.} \right) \ \epsilon^{\mu\nu\rho\sigma}  G^a_{\mu \nu} W^\alpha_{\rho \sigma} \,.
\eea
Matching leads to the following effective couplings:
\begin{equation}
\kappa_g = c_g  \frac{v}{f_\Phi}\,, \quad \kappa_\gamma = ( c_B  + c_W) \frac{v}{2 f_\Phi}\,, \quad \kappa_Z = \left(c_B  - \frac{c_W}{\tan^2 \theta_W} \right) \frac{v}{2 f_\Phi}\,.
\end{equation}

\subsubsection*{Isospin 3}

The minimal choice is to embed the neutral pseudo-scalar in an isotriplet with zero hypercharge, together with charged components:
\begin{equation}
\Phi_3 = \begin{pmatrix} \Phi_3^+ \\ \Phi_{3,0} \\ \Phi_3^- \end{pmatrix}\,.
\end{equation}

The leading couplings arise at dim-5 in the form:
\begin{equation}
\mathcal{L}_{\Phi_3} =  i\ \Phi^{a \alpha}_3\ \left(\frac{\lambda_t}{f_{\Phi}} (\varphi_{H}^\dagger\tau^\alpha \bar{q}_L) \frac{\lambda^a}{2} t_R - \mbox{h.c.} \right)
+ \frac{g_s g \kappa_{W}}{32 \pi^2 f_{\Phi}} \Phi^{a\alpha}_3\ \epsilon^{\mu\nu\rho\sigma}\ G^a_{\mu\nu}W^\alpha_{\rho\sigma}\,.
\end{equation}
In this case, the matching leads to
\begin{equation}
C_t = \frac{\lambda_t v}{2 \sqrt{2} m_t}\,, \quad \kappa_\gamma = \kappa_W\,, \quad \kappa_Z = - \frac{1}{\tan^2 \theta_W} \kappa_W\,.
\end{equation}
Interestingly, in this case couplings to two gluons are suppressed by two Higgs insertions, i.e. they arise at dim-7 level with $\kappa_g \propto v^2/f_\Phi^2$. Thus, decays into $g\gamma$ and $g Z$ are dominant, with 
\begin{equation}
\frac{\mbox{BR} (\Phi_{3,0} \to g Z)}{\mbox{BR} (\Phi_{3,0} \to g \gamma)} \approx \tan^{-2} \theta_W \approx 3.5\,,
\end{equation}
thus dominant decays into $Z$ final states.

\section{Current limits from single production} \label{app:bounds}

	\begin{figure}[t]
	\begin{center}
		\begin{tabular}{cc}
			\includegraphics[width=7.2cm]{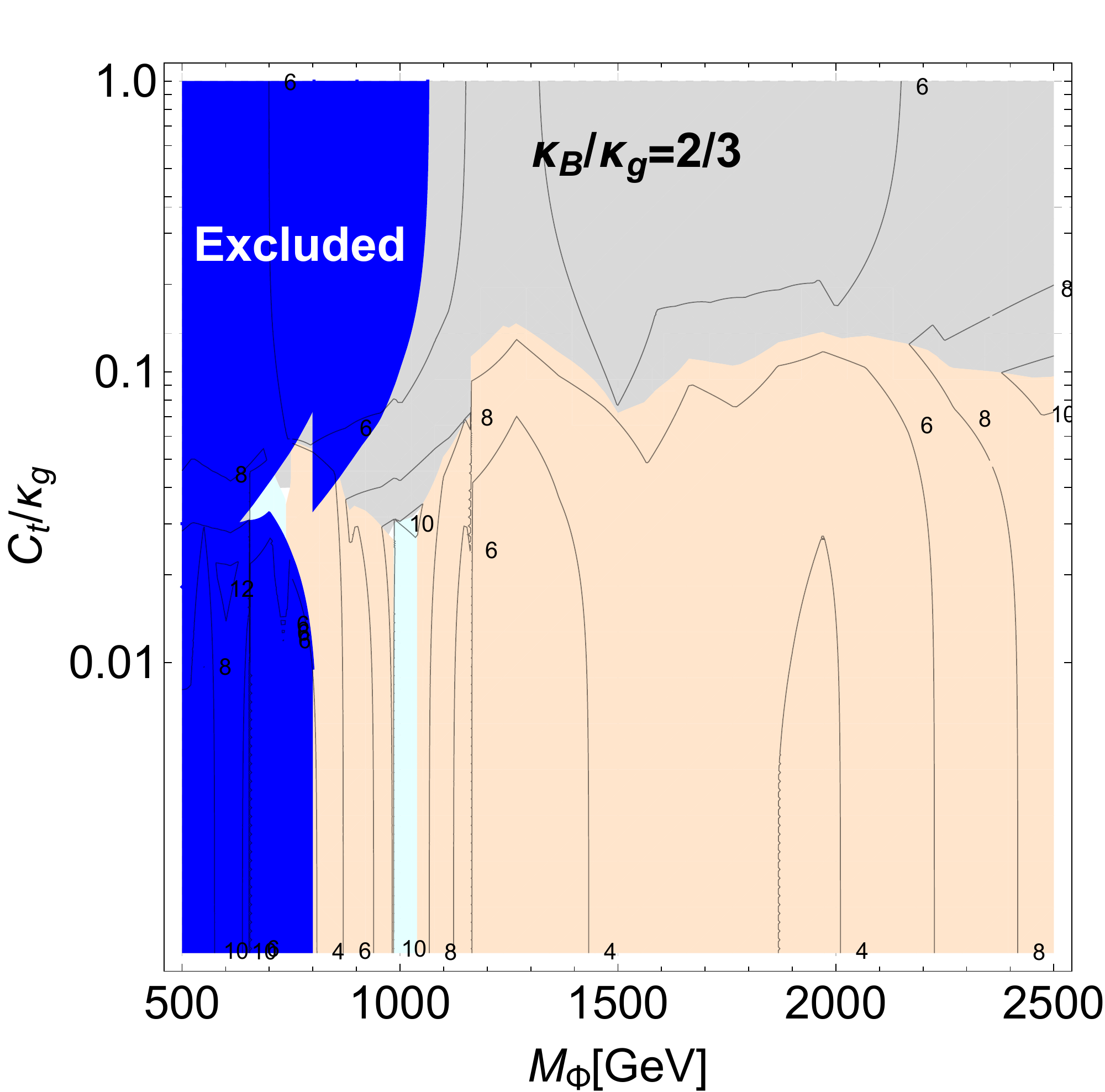}&\includegraphics[width=7.2cm]{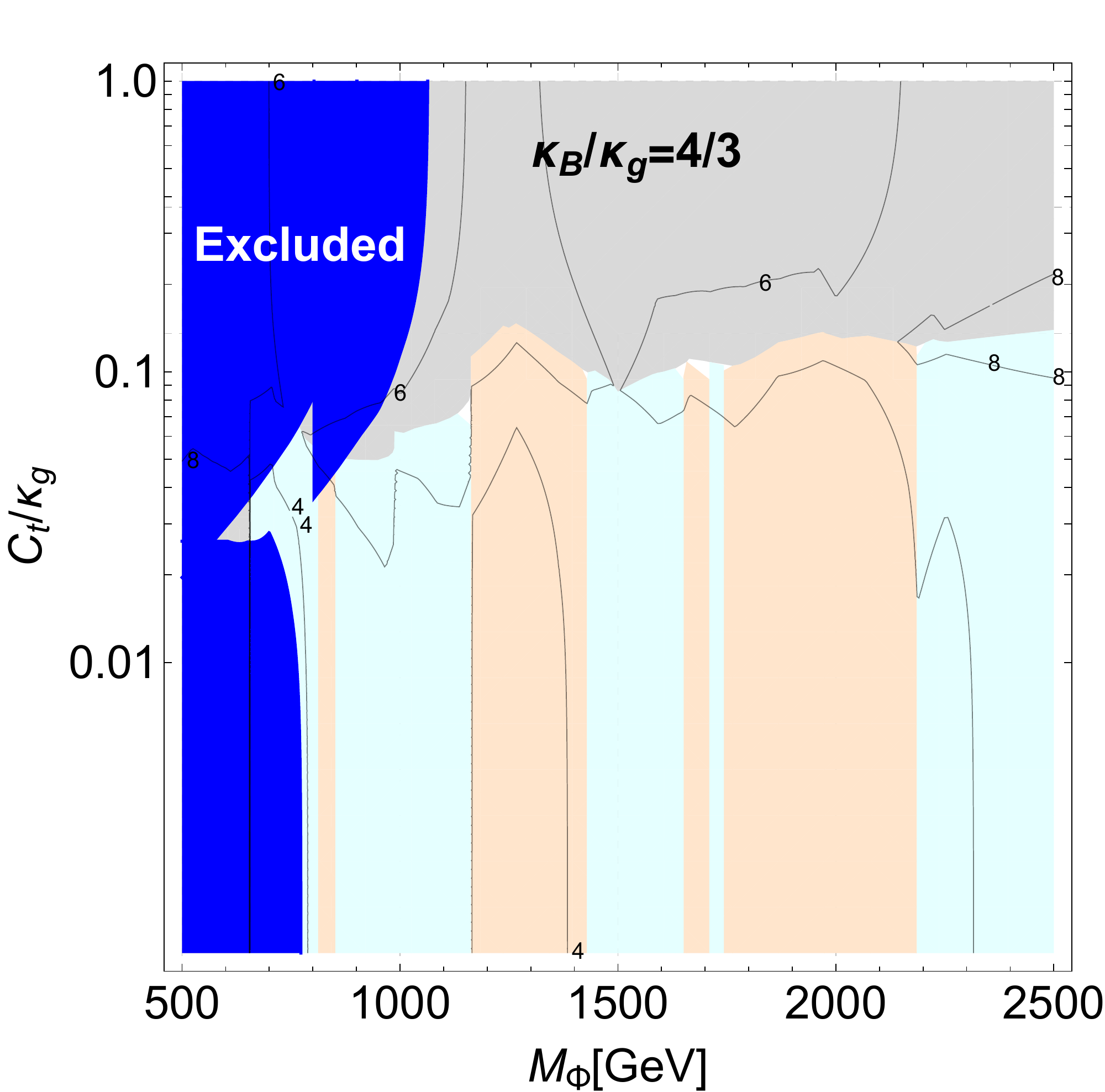}		
		\end{tabular}
	\end{center}
	\caption{Observed bounds on color octet single production in the $M_\Phi$ vs. $C_t/\kappa_g$ plane for  $\kappa_B/\kappa_g = 2/3$ and $4/3$. The contours show the  upper bound on $\kappa_g/f_\Phi$ (in TeV$^{-1}$). The dark-blue areas are excluded by pair production searches. In the  grey areas, the strongest bound arises from $t\bar{t}$ resonance searches. In the orange (cyan) areas, the currently strongest bound arises from di-jet (jet-$\gamma$) searches.}
	\label{fig:singlepsbds2_obs}
\end{figure}

	\begin{figure}[t]
	\begin{center}
		\begin{tabular}{c}
			\includegraphics[width=7.2cm]{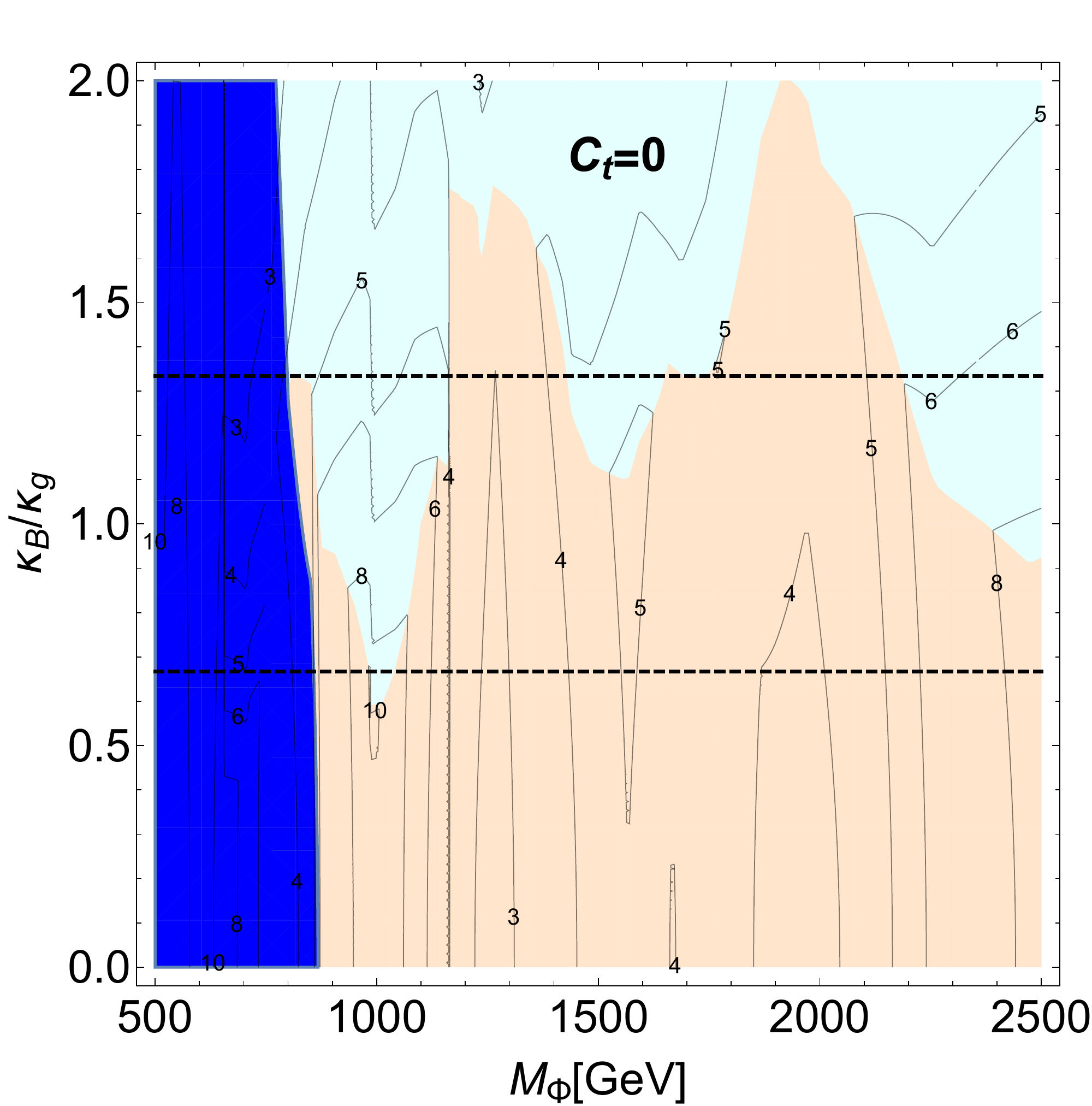}		
		\end{tabular}
	\end{center}
	\caption{Observed bounds on color octet single production in the $M_\Phi$ vs. $\kappa_B/\kappa_g$ plane. The dark blue area is excluded from pair production bounds, while the orange (cyan) area labels where the dominant bound arises from the $jj$ ($j\gamma$) search. 
		Contours show the upper bound on $\kappa_g/f_\Phi$ in TeV$^{-1}$. For reference, the horizontal dashed lines indicate fixed $\kappa_B/\kappa_g$ ratios of $2/3$ (lower) and $4/3$ (higher). }
	\label{fig:singlepsbdsYchi_obs}
\end{figure}

In Section~\ref{sec:photon}, we have presented upper bounds on $\kappa_g/f_\Phi$ deriving from expected bounds from searches sensitive to single production. In this Appendix we present similar plots, drawn from the observed bounds. Fig.~\ref{fig:singlepsbds2_obs} corresponds to Fig.~\ref{fig:singlepsbds2} showing results for the two benchmark models, while Fig.~\ref{fig:singlepsbdsYchi_obs} corresponds to Fig.~\ref{fig:singlepsbdsYchi} in showing limits for $C_t=0$.

Comparing the two pairs of figures clearly highlights that the limits on the coupling $\kappa_g/f_\Phi$ are comparable, as well as the regions of dominance of each final state. The main difference is that the plots in this Appendix are much more irregular, due to the statistical fluctuations in the observed bounds (C.f. Fig.~\ref{fig:singlepsbds}).

\newpage
\bibliographystyle{JHEP-2-2}
\bibliography{biblio}

\providecommand{\href}[2]{#2}\begingroup\raggedright\begin{thebibliography}{10}

\bibitem{Chen:2014haa}
C.-Y. Chen, A.~Freitas, T.~Han and K.~S.~M. Lee, ``{Heavy Color-Octet Particles
  at the LHC},''\href{http://dx.doi.org/10.1007/JHEP05(2015)135}{\emph{JHEP}
  {\bf 05} (2015) 135}, [\href{https://arxiv.org/abs/1410.8113}{{\tt
  1410.8113}}].

\bibitem{Cacciapaglia:2015eqa}
G.~Cacciapaglia, H.~Cai, A.~Deandrea, T.~Flacke, S.~J. Lee and A.~Parolini,
  ``{Composite scalars at the LHC: the Higgs, the Sextet and the
  Octet},''\href{http://dx.doi.org/10.1007/JHEP11(2015)201}{\emph{JHEP} {\bf
  11} (2015) 201}, [\href{https://arxiv.org/abs/1507.02283}{{\tt 1507.02283}}].

\bibitem{Barnard:2013zea}
J.~Barnard, T.~Gherghetta and T.~S. Ray, ``{UV descriptions of composite Higgs
  models without elementary
  scalars},''\href{http://dx.doi.org/10.1007/JHEP02(2014)002}{\emph{JHEP} {\bf
  02} (2014) 002}, [\href{https://arxiv.org/abs/1311.6562}{{\tt 1311.6562}}].

\bibitem{Ferretti:2013kya}
G.~Ferretti and D.~Karateev, ``{Fermionic UV completions of Composite Higgs
  models},''\href{http://dx.doi.org/10.1007/JHEP03(2014)077}{\emph{JHEP} {\bf
  03} (2014) 077}, [\href{https://arxiv.org/abs/1312.5330}{{\tt 1312.5330}}].

\bibitem{Englert:2017bme}
C.~Englert, G.~Ferretti and M.~Spannowsky, ``{Jet-associated resonance
  spectroscopy},''\href{http://dx.doi.org/10.1140/epjc/s10052-017-5416-2}{\emph{Eur.
  Phys. J.} {\bf C77} (2017) 842},
  [\href{https://arxiv.org/abs/1706.04242}{{\tt 1706.04242}}].

\bibitem{Belyaev:2016ftv}
A.~Belyaev, G.~Cacciapaglia, H.~Cai, G.~Ferretti, T.~Flacke, A.~Parolini
  et~al., ``{Di-boson signatures as Standard Candles for Partial
  Compositeness},''\href{http://dx.doi.org/10.1007/JHEP01(2017)094,
  10.1007/JHEP12(2017)088}{\emph{JHEP} {\bf 01} (2017) 094},
  [\href{https://arxiv.org/abs/1610.06591}{{\tt 1610.06591}}].

\bibitem{Hayot:1980gg}
F.~Hayot and O.~Napoly, ``{Detecting a Heavy Colored Object at the Fnal
  Tevatron},''\href{http://dx.doi.org/10.1007/BF01436311}{\emph{Z. Phys.} {\bf
  C7} (1981) 229}.

\bibitem{Belyaev:1999xe}
A.~Belyaev, R.~Rosenfeld and A.~R. Zerwekh, ``{Tevatron potential for
  technicolor search with prompt
  photons},''\href{http://dx.doi.org/10.1016/S0370-2693(99)00869-2}{\emph{Phys.
  Lett.} {\bf B462} (1999) 150--157},
  [\href{https://arxiv.org/abs/hep-ph/9905468}{{\tt hep-ph/9905468}}].

\bibitem{Aad:2015gdg}
{\scshape ATLAS} collaboration, G.~Aad et~al., ``{Analysis of events with
  $b$-jets and a pair of leptons of the same charge in $pp$ collisions at
  $\sqrt{s}=8$ TeV with the ATLAS
  detector},''\href{http://dx.doi.org/10.1007/JHEP10(2015)150}{\emph{JHEP} {\bf
  10} (2015) 150}, [\href{https://arxiv.org/abs/1504.04605}{{\tt 1504.04605}}].

\bibitem{Aad:2015kqa}
{\scshape ATLAS} collaboration, G.~Aad et~al., ``{Search for production of
  vector-like quark pairs and of four top quarks in the lepton-plus-jets final
  state in $pp$ collisions at $\sqrt{s}=8$ TeV with the ATLAS
  detector},''\href{http://dx.doi.org/10.1007/JHEP08(2015)105}{\emph{JHEP} {\bf
  08} (2015) 105}, [\href{https://arxiv.org/abs/1505.04306}{{\tt 1505.04306}}].

\bibitem{Khachatryan:2014lpa}
{\scshape CMS} collaboration, V.~Khachatryan et~al., ``{Search for
  pair-produced resonances decaying to jet pairs in proton–proton collisions
  at $\sqrt{s} =$ 8
  TeV},''\href{http://dx.doi.org/10.1016/j.physletb.2015.04.045}{\emph{Phys.
  Lett.} {\bf B747} (2015) 98--119},
  [\href{https://arxiv.org/abs/1412.7706}{{\tt 1412.7706}}].

\bibitem{Degrande:2014sta}
C.~Degrande, B.~Fuks, V.~Hirschi, J.~Proudom and H.-S. Shao, ``{Automated
  next-to-leading order predictions for new physics at the LHC: the case of
  colored scalar pair
  production},''\href{http://dx.doi.org/10.1103/PhysRevD.91.094005}{\emph{Phys.
  Rev.} {\bf D91} (2015) 094005}, [\href{https://arxiv.org/abs/1412.5589}{{\tt
  1412.5589}}].

\bibitem{Darme:2018dvz}
L.~Darmé, B.~Fuks and M.~Goodsell, ``{Cornering sgluons with four-top-quark
  events},''\href{http://dx.doi.org/10.1016/j.physletb.2018.08.001}{\emph{Phys.
  Lett.} {\bf B784} (2018) 223--228},
  [\href{https://arxiv.org/abs/1805.10835}{{\tt 1805.10835}}].

\bibitem{Sirunyan:2017roi}
{\scshape CMS} collaboration, A.~M. Sirunyan et~al., ``{Search for standard
  model production of four top quarks with same-sign and multilepton final
  states in proton–proton collisions at $\sqrt{s} = 13\,\text {TeV}
  $},''\href{http://dx.doi.org/10.1140/epjc/s10052-018-5607-5}{\emph{Eur. Phys.
  J.} {\bf C78} (2018) 140}, [\href{https://arxiv.org/abs/1710.10614}{{\tt
  1710.10614}}].

\bibitem{Sirunyan:2019wxt}
{\scshape CMS} collaboration, A.~M. Sirunyan et~al., ``{Search for production
  of four top quarks in final states with same-sign or multiple leptons in
  proton-proton collisions at $\sqrt{s}=$ 13 TeV},''
  \href{https://arxiv.org/abs/1908.06463}{{\tt 1908.06463}}.

\bibitem{Aaboud:2018xuw}
{\scshape ATLAS} collaboration, M.~Aaboud et~al., ``{Search for pair production
  of up-type vector-like quarks and for four-top-quark events in final states
  with multiple $b$-jets with the ATLAS
  detector},''\href{http://dx.doi.org/10.1007/JHEP07(2018)089}{\emph{JHEP} {\bf
  07} (2018) 089}, [\href{https://arxiv.org/abs/1803.09678}{{\tt 1803.09678}}].

\bibitem{Aaboud:2018xpj}
{\scshape ATLAS} collaboration, M.~Aaboud et~al., ``{Search for new phenomena
  in events with same-charge leptons and $b$-jets in $pp$ collisions at
  $\sqrt{s}= 13$ TeV with the ATLAS
  detector},''\href{http://dx.doi.org/10.1007/JHEP12(2018)039}{\emph{JHEP} {\bf
  12} (2018) 039}, [\href{https://arxiv.org/abs/1807.11883}{{\tt 1807.11883}}].

\bibitem{Aaboud:2018jsj}
{\scshape ATLAS} collaboration, M.~Aaboud et~al., ``{Search for four-top-quark
  production in the single-lepton and opposite-sign dilepton final states in pp
  collisions at $\sqrt{s}$ = 13 TeV with the ATLAS
  detector},''\href{http://dx.doi.org/10.1103/PhysRevD.99.052009}{\emph{Phys.
  Rev.} {\bf D99} (2019) 052009}, [\href{https://arxiv.org/abs/1811.02305}{{\tt
  1811.02305}}].

\bibitem{Sirunyan:2019nxl}
{\scshape CMS} collaboration, A.~M. Sirunyan et~al., ``{Search for the
  production of four top quarks in the single-lepton and opposite-sign dilepton
  final states in proton-proton collisions at $ \sqrt{s} $ = 13
  TeV},''\href{http://dx.doi.org/10.1007/JHEP11(2019)082}{\emph{JHEP} {\bf 11}
  (2019) 082}, [\href{https://arxiv.org/abs/1906.02805}{{\tt 1906.02805}}].

\bibitem{Aaboud:2017nmi}
{\scshape ATLAS} collaboration, M.~Aaboud et~al., ``{A search for pair-produced
  resonances in four-jet final states at $\sqrt{s} =$ 13 TeV with the ATLAS
  detector},''\href{http://dx.doi.org/10.1140/epjc/s10052-018-5693-4}{\emph{Eur.
  Phys. J.} {\bf C78} (2018) 250},
  [\href{https://arxiv.org/abs/1710.07171}{{\tt 1710.07171}}].

\bibitem{Sirunyan:2018rlj}
{\scshape CMS} collaboration, A.~M. Sirunyan et~al., ``{Search for
  pair-produced resonances decaying to quark pairs in proton-proton collisions
  at $\sqrt{s}=$ 13
  TeV},''\href{http://dx.doi.org/10.1103/PhysRevD.98.112014}{\emph{Phys. Rev.}
  {\bf D98} (2018) 112014}, [\href{https://arxiv.org/abs/1808.03124}{{\tt
  1808.03124}}].

\bibitem{Bhattacherjee:2017cxh}
B.~Bhattacherjee, P.~Byakti, A.~Kushwaha and S.~K. Vempati, ``{Unification with
  Vector-like fermions and signals at
  LHC},''\href{http://dx.doi.org/10.1007/JHEP05(2018)090}{\emph{JHEP} {\bf 05}
  (2018) 090}, [\href{https://arxiv.org/abs/1702.06417}{{\tt 1702.06417}}].

\bibitem{Sirunyan:2018ryr}
{\scshape CMS} collaboration, A.~M. Sirunyan et~al., ``{Search for resonant $
  \mathrm{t}\overline{\mathrm{t}} $ production in proton-proton collisions at $
  \sqrt{s}=13 $
  TeV},''\href{http://dx.doi.org/10.1007/JHEP04(2019)031}{\emph{JHEP} {\bf 04}
  (2019) 031}, [\href{https://arxiv.org/abs/1810.05905}{{\tt 1810.05905}}].

\bibitem{Aaboud:2018mjh}
{\scshape ATLAS} collaboration, M.~Aaboud et~al., ``{Search for heavy particles
  decaying into top-quark pairs using lepton-plus-jets events in
  proton–proton collisions at $\sqrt{s} = 13$ $\text {TeV}$ with the ATLAS
  detector},''\href{http://dx.doi.org/10.1140/epjc/s10052-018-5995-6}{\emph{Eur.
  Phys. J.} {\bf C78} (2018) 565},
  [\href{https://arxiv.org/abs/1804.10823}{{\tt 1804.10823}}].

\bibitem{Aaboud:2019roo}
{\scshape ATLAS} collaboration, M.~Aaboud et~al., ``{Search for heavy particles
  decaying into a top-quark pair in the fully hadronic final state in $pp$
  collisions at $\sqrt{s} =$ 13 TeV with the ATLAS
  detector},''\href{http://dx.doi.org/10.1103/PhysRevD.99.092004}{\emph{Phys.
  Rev.} {\bf D99} (2019) 092004}, [\href{https://arxiv.org/abs/1902.10077}{{\tt
  1902.10077}}].

\bibitem{Sirunyan:2018xlo}
{\scshape CMS} collaboration, A.~M. Sirunyan et~al., ``{Search for narrow and
  broad dijet resonances in proton-proton collisions at $ \sqrt{s}=13 $ TeV and
  constraints on dark matter mediators and other new
  particles},''\href{http://dx.doi.org/10.1007/JHEP08(2018)130}{\emph{JHEP}
  {\bf 08} (2018) 130}, [\href{https://arxiv.org/abs/1806.00843}{{\tt
  1806.00843}}].

\bibitem{CMS:2018wxx}
{\scshape CMS} collaboration, ``{Searches for dijet resonances in pp collisions
  at $\sqrt{s}=13~\mathrm{TeV}$ using the 2016 and 2017 datasets},''
  CMS-PAS-EXO-17-026.

\bibitem{Aaboud:2017yvp}
{\scshape ATLAS} collaboration, M.~Aaboud et~al., ``{Search for new phenomena
  in dijet events using 37 fb$^{-1}$ of $pp$ collision data collected at
  $\sqrt{s}=$13 TeV with the ATLAS
  detector},''\href{http://dx.doi.org/10.1103/PhysRevD.96.052004}{\emph{Phys.
  Rev.} {\bf D96} (2017) 052004}, [\href{https://arxiv.org/abs/1703.09127}{{\tt
  1703.09127}}].

\bibitem{Aaboud:2018fzt}
{\scshape ATLAS} collaboration, M.~Aaboud et~al., ``{Search for low-mass dijet
  resonances using trigger-level jets with the ATLAS detector in $pp$
  collisions at $\sqrt{s}=13$
  TeV},''\href{http://dx.doi.org/10.1103/PhysRevLett.121.081801}{\emph{Phys.
  Rev. Lett.} {\bf 121} (2018) 081801},
  [\href{https://arxiv.org/abs/1804.03496}{{\tt 1804.03496}}].

\bibitem{Aad:2019hjw}
{\scshape ATLAS} collaboration, G.~Aad et~al., ``{Search for new resonances in
  mass distributions of jet pairs using 139 fb$^{-1}$ of $pp$ collisions at
  $\sqrt{s}=13$ TeV with the ATLAS detector},''
  \href{https://arxiv.org/abs/1910.08447}{{\tt 1910.08447}}.

\bibitem{Aaboud:2017nak}
{\scshape ATLAS} collaboration, M.~Aaboud et~al., ``{Search for new phenomena
  in high-mass final states with a photon and a jet from $pp$ collisions at
  $\sqrt{s}$ = 13 TeV with the ATLAS
  detector},''\href{http://dx.doi.org/10.1140/epjc/s10052-018-5553-2}{\emph{Eur.
  Phys. J.} {\bf C78} (2018) 102},
  [\href{https://arxiv.org/abs/1709.10440}{{\tt 1709.10440}}].

\bibitem{Sirunyan:2017fho}
{\scshape CMS} collaboration, A.~M. Sirunyan et~al., ``{Search for excited
  quarks of light and heavy flavor in $\gamma+$jet final states in
  proton-proton collisions as $\sqrt{s}=$13
  TeV},''\href{http://dx.doi.org/10.1016/j.physletb.2018.04.007}{\emph{Phys.
  Lett.} {\bf B781} (2018) 390--411},
  [\href{https://arxiv.org/abs/1711.04652}{{\tt 1711.04652}}].

\bibitem{Aad:2013cva}
{\scshape ATLAS} collaboration, G.~Aad et~al., ``{Search for new phenomena in
  photon+jet events collected in proton--proton collisions at sqrt(s) = 8 TeV
  with the ATLAS
  detector},''\href{http://dx.doi.org/10.1016/j.physletb.2013.12.029}{\emph{Phys.
  Lett.} {\bf B728} (2014) 562--578},
  [\href{https://arxiv.org/abs/1309.3230}{{\tt 1309.3230}}].

\bibitem{Khachatryan:2014aka}
{\scshape CMS} collaboration, V.~Khachatryan et~al., ``{Search for excited
  quarks in the $\gamma +$jet final state in proton–proton collisions at
  $\sqrt s=8$
  TeV},''\href{http://dx.doi.org/10.1016/j.physletb.2014.09.048}{\emph{Phys.
  Lett.} {\bf B738} (2014) 274--293},
  [\href{https://arxiv.org/abs/1406.5171}{{\tt 1406.5171}}].

\bibitem{Cacciapaglia:2020vyf}
G.~Cacciapaglia, A.~Deandrea, T.~Flacke and A.~M. Iyer, ``{Gluon-Photon
  Signatures for color octet at the LHC (and beyond)},''
  \href{https://arxiv.org/abs/2002.01474}{{\tt 2002.01474}}.

\bibitem{Bizot:2018tds}
N.~Bizot, G.~Cacciapaglia and T.~Flacke, ``{Common exotic decays of top
  partners},''\href{http://dx.doi.org/10.1007/JHEP06(2018)065}{\emph{JHEP} {\bf
  06} (2018) 065}, [\href{https://arxiv.org/abs/1803.00021}{{\tt 1803.00021}}].

\bibitem{Sjostrand:2007gs}
T.~Sjostrand, S.~Mrenna and P.~Z. Skands, ``{A Brief Introduction to PYTHIA
  8.1},''\href{http://dx.doi.org/10.1016/j.cpc.2008.01.036}{\emph{Comput. Phys.
  Commun.} {\bf 178} (2008) 852--867},
  [\href{https://arxiv.org/abs/0710.3820}{{\tt 0710.3820}}].

\bibitem{deFavereau:2013fsa}
{\scshape DELPHES 3} collaboration, J.~de~Favereau, C.~Delaere, P.~Demin,
  A.~Giammanco, V.~Lemaître, A.~Mertens et~al., ``{DELPHES 3, A modular
  framework for fast simulation of a generic collider
  experiment},''\href{http://dx.doi.org/10.1007/JHEP02(2014)057}{\emph{JHEP}
  {\bf 02} (2014) 057}, [\href{https://arxiv.org/abs/1307.6346}{{\tt
  1307.6346}}].

\bibitem{CMS:twa}
{\scshape CMS} collaboration, ``{Measurement of triple-differential cross
  section of gamma+jet production},'' CMS-PAS-QCD-11-005.

\bibitem{Alwall:2014hca}
J.~Alwall, R.~Frederix, S.~Frixione, V.~Hirschi, F.~Maltoni, O.~Mattelaer
  et~al., ``{The automated computation of tree-level and next-to-leading order
  differential cross sections, and their matching to parton shower
  simulations},''\href{http://dx.doi.org/10.1007/JHEP07(2014)079}{\emph{JHEP}
  {\bf 07} (2014) 079}, [\href{https://arxiv.org/abs/1405.0301}{{\tt
  1405.0301}}].

\bibitem{Cowan:2010js}
G.~Cowan, K.~Cranmer, E.~Gross and O.~Vitells, ``{Asymptotic formulae for
  likelihood-based tests of new
  physics},''\href{http://dx.doi.org/10.1140/epjc/s10052-011-1554-0,
  10.1140/epjc/s10052-013-2501-z}{\emph{Eur. Phys. J.} {\bf C71} (2011) 1554},
  [\href{https://arxiv.org/abs/1007.1727}{{\tt 1007.1727}}].

\bibitem{Manohar:2006ga}
A.~V. Manohar and M.~B. Wise, ``{Flavor changing neutral currents, an extended
  scalar sector, and the Higgs production rate at the CERN
  LHC},''\href{http://dx.doi.org/10.1103/PhysRevD.74.035009}{\emph{Phys. Rev.}
  {\bf D74} (2006) 035009}, [\href{https://arxiv.org/abs/hep-ph/0606172}{{\tt
  hep-ph/0606172}}].

\bibitem{Arnold:2009ay}
J.~M. Arnold, M.~Pospelov, M.~Trott and M.~B. Wise, ``{Scalar Representations
  and Minimal Flavor
  Violation},''\href{http://dx.doi.org/10.1007/JHEP01(2010)073}{\emph{JHEP}
  {\bf 01} (2010) 073}, [\href{https://arxiv.org/abs/0911.2225}{{\tt
  0911.2225}}].

\end{thebibliography}\endgroup

\end{document}